\documentclass[a4paper,fleqn]{cas-sc}

\usepackage[numbers]{natbib}
\usepackage{lineno}
\usepackage{textcomp}
\usepackage{anyfontsize}
\usepackage{graphicx}
\usepackage{graphics}
\usepackage{geometry}
\usepackage{adjustbox,lipsum}
\usepackage{float}
\usepackage{amssymb}
\usepackage{color}
\usepackage{xcolor}  
\usepackage{hyperref}  
\hypersetup{colorlinks,urlcolor=blue,linkcolor=blue,citecolor=blue,filecolor=blue,urlcolor=blue}
\usepackage{multirow}
\usepackage{booktabs}
\usepackage{longtable}
\usepackage{lscape}
\usepackage[final]{pdfpages} 

\DeclareMathOperator*{\argmax}{argmax}

\begin{document}
	\let\WriteBookmarks\relax
	\def\floatpagepagefraction{1}
	\def\textpagefraction{.001}
    \shorttitle{Validation of Newcomb-Benford law and Weibull distribution at NMJ}
     \shortauthors{A J da Silva et~al.}


\title[mode = title]{On the validation of Newcomb-Benford law and Weibull distribution in neuromuscular transmission}                      

\author[1]{A J da Silva}
\cormark[1]
\ead{adjesbr@ufsb.edu.br,adjesbr@gmail.com}
\address[1]{Instituto de Humanidades, Artes e Ci\^encias, Universidade Federal do Sul da Bahia, $45613-204$, Itabuna, Bahia.  Brazil}
\credit{Conceptualization of this study, Experimental measurements, Data analysis, Original draft preparation, Writing}

\author[2]{S Floquet}
\address[2]{Colegiado de Engenharia Civil, Universidade Federal do Vale do S\~ao Francisco, $48902-300$, Juazeiro, Bahia, Brazil}
\credit{Data analysis, Original draft preparation}

\author[3]{D O C Santos}
\credit{Data analysis, Original draft preparation}

\author[4]{R F Lima}
\address[3]{Departamento de Fisiologia e Farmacologia, Faculdade de Medicina, Universidade Federal do Cear\'a, $60430-270$, Fortaleza, Cear\'a, Brazil}
\credit{Experimental measurements}

\cortext[cor1]{Corresponding author}

\begin{abstract}
The neuromuscular junction represents a relevant substrate for revealing important biophysical mechanisms of synaptic transmission. In this context, calcium ions are important in the synapse machinery, providing the nervous impulse transmission to the muscle fiber. In this work, we carefully investigated whether intervals of spontaneous electrical activity, recorded in seven different calcium concentrations, conform Newcomb-Benford law. Our analysis revealed that electrical discharge of neuromuscular junction obeys the expected values for Newcomb-Benford law for first and second digits, while first-two digits do not perfectly follows the law. We next examined previous theoretical studies, establishing  a relation between the law and lognormal and Weibull distributions. We showed that Weibull distribution is more appropriate to fit the intervals as compared to lognormal distribution. Altogether, the present findings strongly suggest that spontaneous activity is a base-scale invariant phenomenon.  
\end{abstract}


\maketitle  

\section{Introduction}

The neuromuscular junction (NMJ) is responsible for communicating electrical impulses from the motor neuron to the skeletal muscle, yielding muscular contraction \cite{sanes}. The terminal formed by the NMJ constitute a well-studied case of chemical synapse. Facility of tissue extraction and stereotyped electrical response of nervous activity, represent some of the advantages of using NMJ in biophysical research. 

Bernard Katz led most of the pioneering work on the mechanistic basis of neuromuscular transmission \cite{katz1}. Making use of electrophysiological recordings, Katz and Fatt discovered spontaneous small subthreshold depolarization, called miniature end-plate potential (MEPP) \cite{katz2}. These peculiar signals were further interpreted as due to a single vesicle fusion with the membrane terminal, configuring the vesicular hypothesis. Most importantly, NMJ emerged as an attractive substrate for combining mathematical modelling with empirical protocols. This enabled to elucidate several mechanisms involved in neurotransmitter transmission. For instance, studies revealed that MEPPs are no longer constant in size or temporal distribution. These investigations associated MEPPs occurrence as governed by Gaussian and Poisson statistics \cite{bennett}. Curiously, Katz already had attempted to point out weaknesses in the Poissonian predictions. In fact, Poisson and Gaussian models, commonly used to access the quantal nature of neurosecretion, require uniformity and stationarity assumptions \cite{Higashima1990}. However, to take into account such noticeable deviation from the early studies, a more general mathematical structure for explaining MEPP firing dynamics was developed. This approach allowed to reveal scale invariance or fractality embedded in several NMJ preparations \cite{washio,lowen,takeda}. 

In a previous work, we showed the existence of long-range correlations associated to MEPP discharge at the NMJ of mouse diaphragms  \cite{adjesbr}. We demonstrated that \textit{q}-Gaussian distributions are more accurate in describing MEPP amplitudes as compared to the Gaussian function. These results suggest that spontaneous secretion of neurotransmitters exhibits scale invariance and long-range correlations. Furthermore, in two independent studies, Robinson and Van der Kloot used Gamma, Weibull and lognormal functions to examine MEPP amplitudes \cite{kloot1989,robinson1976}. In addition, two reports analysing chemical synapses from the brain, concluded that quantal statistics can be better understood by presuming Pascal and Weibull distributions, respectively \cite{Higashima1990,Kamiya1992,McKeegan2002}. On the other hand, other authors also advocate that the lognormal distribution is appropriate to study interbursts harvested from brain synapses \cite{buzsaki,Levine1991}. However, to our knowledge, Weibull statistics has not been applied to analyse MEPP intervals.  

In living organisms, ions are responsible for mediating several physiological functions. Among many ionic species, one can highlight potassium, which is responsible for the resting membrane potential. In addition, sodium is crucial for triggering the action potential generation, while calcium ions $(Ca^{2+})$ deserve attention due to their vital function in triggering the neurotransmission, also acting as a second messenger in molecular signalling \cite{rusakov2006}. Since the discovery of MEPPs, systematic research has revealed how extracellular calcium concentration ($[Ca^{2+}]_{o}$) modify their frequency. Yet, after invade the nerve terminal $Ca^{2+}$ from extracellular ambient interacts with proteins within the synaptic terminal that are responsible for exocytosis. Manipulation of $[Ca^{2+}]_{o}$ also allows the modulation of the degree of neuronal plasticity as a function of vesicular dynamics. Thus, the experimental protocol assuming $[Ca^{2+}]_{o}$ modulation is a suitable manner to verify if machinery behind the MEPP time series follows the first digit phenomenon, also to study if Weibull or longonormal distributions are a more appropriate approach to adjust the MEPP intvervals. 

The anomalous numbers law was pioneered and documented by the polymath Simon Newcomb in 1881, but only after 57 years were his observations revisited and popularised thanks to Frank Benford, who analysed different types of data \cite{newcomb,benford}. This intriguing numerical phenomenon is now known as Newcomb-Benford law (NBL). This counterintuitive law is among the several power or scaling laws found in many physical systems. In mathematical terms, the probabilities for occurrences of first digit are inferred by: 

\begin{eqnarray}
 P( D_{1}  = d_{1} ) & = & \log \left(1 + \frac{1}{d_{1}}  \right), \qquad  d_{1} \in \left\lbrace 1,2,...,9 \right\rbrace \label{benford}
\end{eqnarray}
 
It is important to mention that second digit analysis has also been shown to be relevant in NBL validation. For example, Diekmann successfully identified that articles published in the American Journal of Sociology are well described by taking the second digit into account \cite{Diekmann2007}. Also, Mebane argued that this digit is relevant for detection of election frauds. Probabilities for the appearance of a second digit are quantified by:

\begin{eqnarray}
 P(  D_{2} = d_{2} ) & = & \sum_{d_{1}=1}^{9} \log \left( 1+  \frac{1}{d_{1}d_{2}}  \right), \qquad d_{2} \in \left\lbrace 0,1,...,9 \right\rbrace \label{benford-2}
\end{eqnarray}

Although not frequently used, first-two digits analysis was applied by Nigrini and Miller to investigate accounts payable data and hydrometric statistics \cite{Miller2009,miller2007}. Its functional form is presented below:

\begin{eqnarray}
 P( D_{1}D_{2} = d_{1}d_{2} ) & = & \log \left( 1 + \frac{1}{d_{1}d_{2}}  \right), \qquad d_{1}d_{2} \in \left\lbrace 10,11,...,99 \right\rbrace \label{benford-2b}
\end{eqnarray}

Beyond the heuristic formulation, rigorous mathematical descriptions have been proposed to explain why certain data conforms to the first digit phenomenon. For instance, Pinkham argued that if NBL obeys some universal distribution, then this law should be scale invariant to the units chosen \cite{pinkham}. In Physiology, relationships that depend on spatial invariance can have profound morphological implications, being documented in heart, lung and brain \cite{stanley2001,fadel2004,cai2007,gisiger2000}. In these schemes, if a certain data follows NBL, it must exhibit a base invariance profile \cite{hill95}. Finally, Hill contributed with a rigorous statistical pillar, inserting the law as a branch of modern probability theory, showing that NBL is related to scale invariance \cite{hill95}. In this sense, investigations have examined how closely a dynamical system fulfils NBL and how this law is remarkable for distinguishing noise from chaos \cite{li2015,Snyder2001}. 

Subsequent studies also showed a relationship between NBL and lognormal and Weibull distributions. According to Rodriguez, given certain conditions, datasets that conform to NBL also exhibit adjustments with a lognormal distribution \cite{Rodriguez2004}. On the other hand, Cuff \textit{et al.} pointed out that given particular conditions, NBL is close to the Weibull distribution \cite{millerweibullCuff2015}. Motivated by these authors and by the lack of studies examining the statistics of MEPP intervals, we decide to assume both functions in our work. Among many confirmations in different fields, NBL was verified in physical constants, number of cells in colonies of the cyanobacterium, alpha decay half-lives, and fraud-detection \cite{burke,costas,buck,nigrini}. It is important mention that NBL is the only distribution that is not derived from stationary processes. This observation is particularly suitable in time series analysis from physiological data \cite{li2015}. In Physiology, NBL confirmation involves dynamical transitions in cardiac models and brain electrical activity \cite{sinha2015,kreuzer2014}. Nevertheless, even though this law has been attested in different biological systems, it remains to be verified at the synaptic level. Thus, the primary goal of this work is to investigate if MEPP time series follows the numerical predictions established by NBL. To rigorously address this issue, physiological $[Ca^{2+}]_{o}$, such as values above and below the physiological level, will be assumed in order to modulate the MEPP frequency. In this sense, this strategy also gives rise to the opportunity: (a) assess if NBL is valid only in the physiological concentration of $[Ca^{2+}]_{o}$ or if it is independent of $[Ca^{2+}]_{o}$; (b) check how robust is the conformity as function of the ionic manipulation; (c) test the hypotheses of other authors, stating that classical NBL is more frequently realized in nature; (d) verify if MEPP time series conform to NBL, then demonstrate which distribution, Weibull or the lognormal, is the most appropriate statistics to investigate the intervals.

\section{Materials and Methods}

\subsection{Experimental data acquisition}

The experimental procedures here adopted were approved by the Animal Research Committee (CEUA - UFC, protocol 130/2017). Roughly speaking, adult mice were euthanized by cervical dislocation, diaphragms were extracted and inserted into a physiological artificial fluid (Ringer solution). The following successive $[Ca^{2+}]_{o}$ were used (mM): 0.6, 1.2, 1.8, 2.4, 4.8, 10 and 15. The external composition for $[Ca^{2+}]_{o} = (0.6-4.8)$ mM contained the following concentrations (mM): NaCl (137), NaHCO$_{3}$ (26), KCl (5), NaH$_{2}$PO$_{4}$ (1.2), glucose (10), and MgCl$_{2}$ (1.3). The external solution for $[Ca^{2+}]_{o} = (10-15)$ mM  contained (mM): NaCl (137), NaHCO$_{3}$ (12), KCl (5), NaH$_{2}$PO$_{4}$ (0.3), glucose (10), and MgCl$_{2}$ (1.3). In both solutions pH was adjusted to 7.4 after gassing with 95\%  O$_{2}$ - 5\% CO$_{2}$. Tissues were left bathing in the solution for 30 minutes before the electrophysiological recordings began, minimizing the mechanic trauma of their extraction. Next, tissues were gently transferred to a recording chamber continuously bathed with artificial solution at $T = 24 \pm 1^\circ{}C$. Standard intracellular recording technique was performed to monitor the frequency of spontaneous MEPP by inserting a micropipette at the chosen muscle fiber. We employed borosilicate glass microelectrodes with resistances of 8 - 15 M$\Omega$ when filled with 3 M KCl in electrophysiological acquisitions. Strathclyde Electrophysiology Software (John Dempster, University of Strathclyde), R Language \cite{rsoftware}, Origin (OriginLab, Northampton, MA), and MATLAB (The MathWorks, Inc., Natick, MA) were employed for electrophysiological acquisition and data analysis. 

\begin{figure}
\centering
\includegraphics[width=8cm]{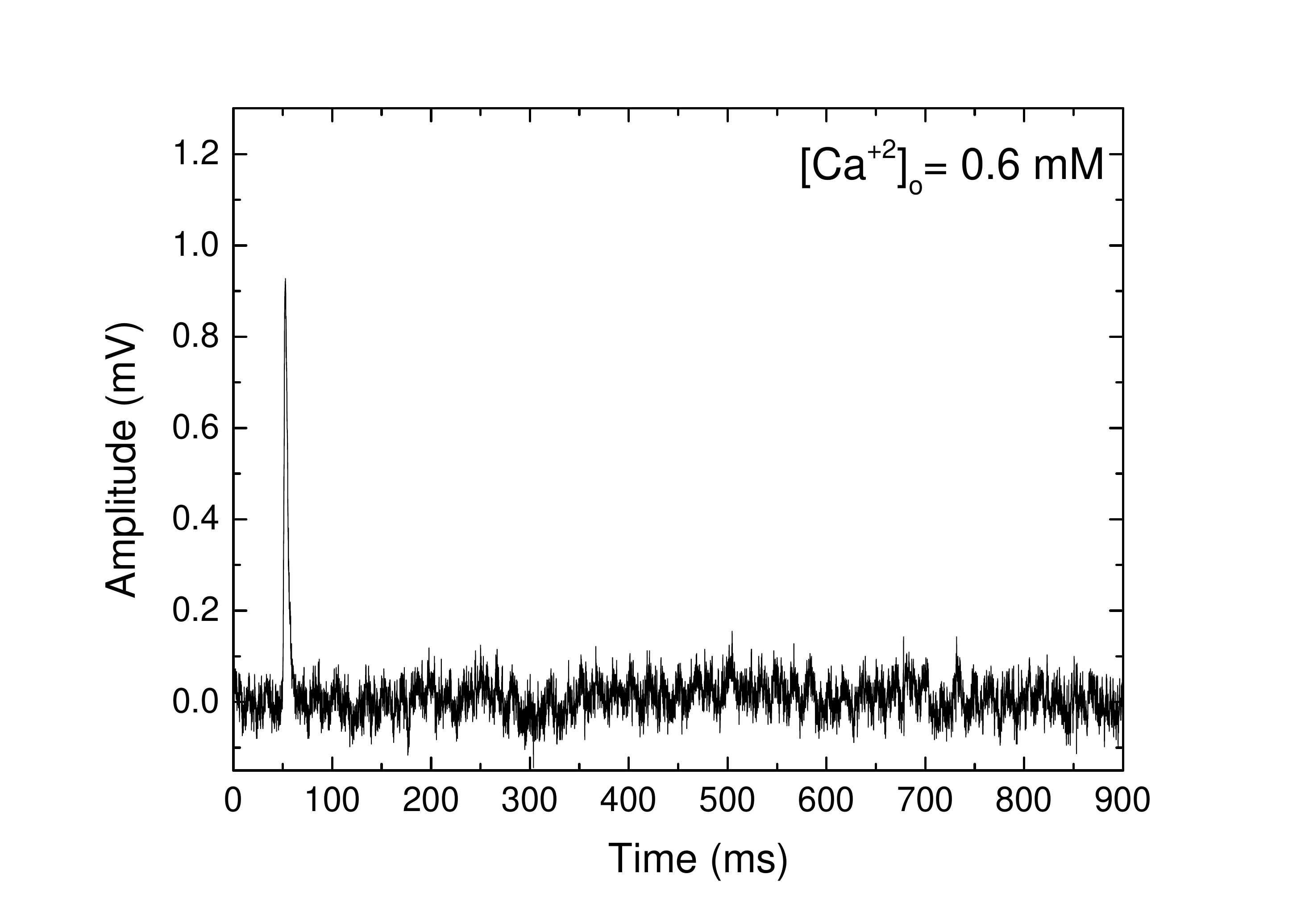}
\vfill
\includegraphics[width=8cm]{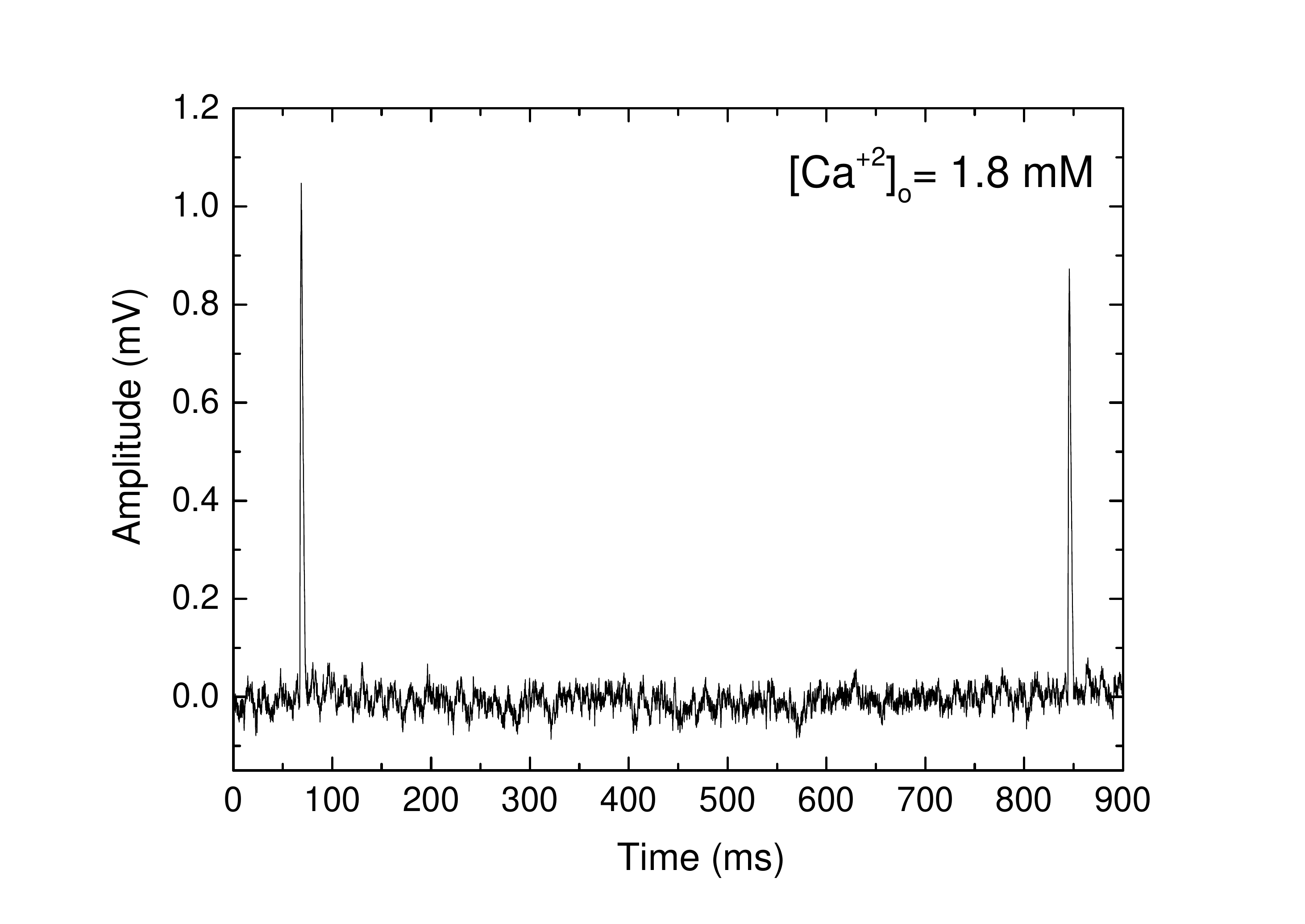}
\vfill
\includegraphics[width=8cm]{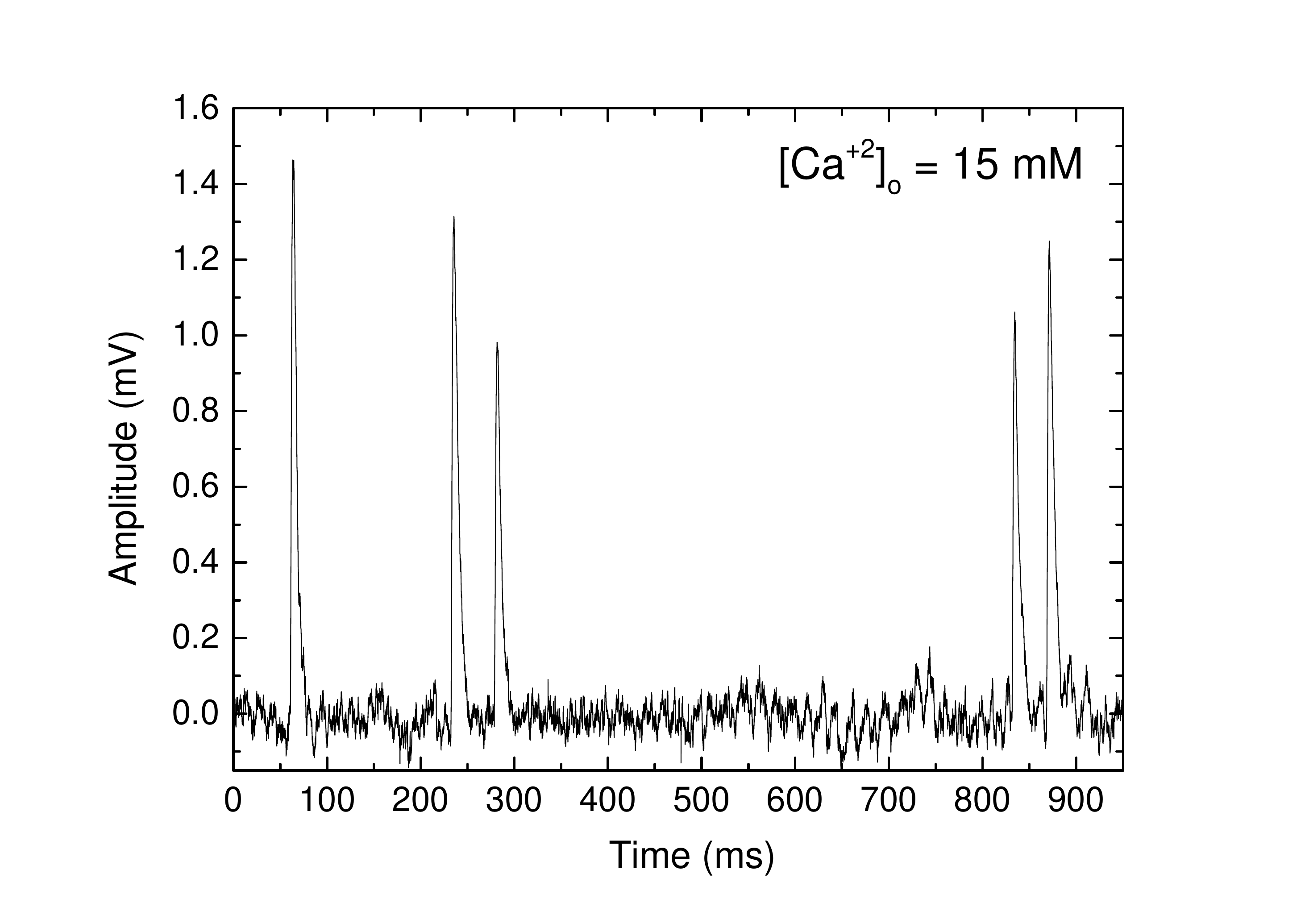}
\caption{Intervals of electrophysiological recordings showing MEPPs for three different ionic concentrations ($[Ca^{2+}]_{o}$ = 0.6, 1.8, 15 mM ).} 
\label{fig1}
\end{figure}

\begin{figure}
\centering
\includegraphics[width=9cm]{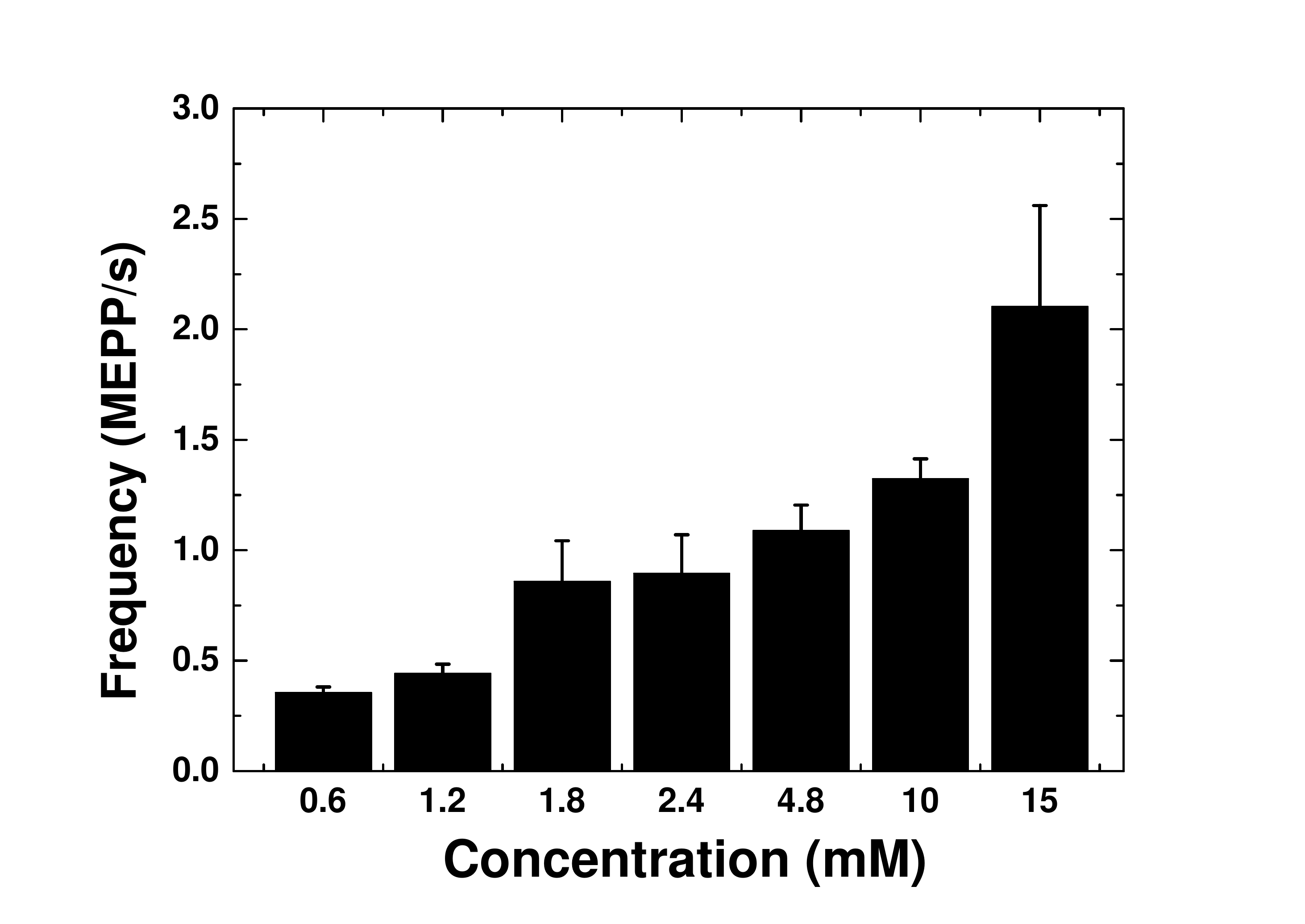}
\caption{Statistical resume showing how $[Ca^{2+}]_{o}$ influences MEPP frequency.}
\label{fig2}
\end{figure}

The experimental paradigm used here afforded a rigorous analysis, where 125,565 MEPP intervals were collected from 78 experiments, with the interval number varying between 213-4123 events. Table \ref{tab1} brings a statistical summary taking into account each concentration. Figure \ref{fig1} shows a representative segment from a typical recording in three $[Ca^{2+}]_{o}$. The statistical resume given by figure \ref{fig2} attests that on average, when $[Ca^{2+}]_{o}$ increases there is a rising in the MEPP discharge rate. In summary,beyond confirming results obtained in other studies, these control experiments guarantee a reliable analysis for all subsequent work \cite{hubbard1961}. 

\begin{table} 
\caption{Electrophysiological parameters obtained in the recordings for each $[Ca^{2+}]_{o}$.}\label{tab1}
\begin{adjustbox}{max width=\textwidth}
\begin{tabular}{@{}ccccc} \toprule

$[Ca^{2+}]_{o}$  & Number of& Number  & Membrane  & Frequency  \\
(mM) & neurojunctions (n) & MEPPS (N) & potential (mV) & ( MEPP/s) \\
\midrule
 $0.6$   &$14$& $1055.14\pm763.88$ &  $59.08\pm9.85$  & $0.35\pm0.10$\\
 $1.2$   &$11$& $1545.82\pm1063.11 $& $-66.26\pm12.31$ & $0.44\pm0.14$\\
 $1.8$   &$14$& $1611.07\pm810.16$   & $-64.15\pm6.34$  &  $0.88\pm0.69$\\
 $2.4$   &$7$&  $777.14\pm242.36$  & $-60.73\pm8.95$  &  $0.89\pm0.46$\\
 $4.8$   &$7$&  $1109.71\pm356.64$  & $-67.10\pm7.26$  &  $1.10\pm0.31$\\
 $10.0$  &$13$& $2394.15\pm668.98$ & $-70.90\pm8.85$  & $1.32\pm0.33$ \\
 $15.0$  &$12$& $2229.00\pm1063.11$ & $-71.44\pm13.85$   &  $2.10\pm1.59$\\

\bottomrule
\end{tabular}
\end{adjustbox}
\end{table}

\subsection{Generalized NBL}

Despite displaying asymmetric distributions, a number of phenomena, apparently do not follow NBL. Examples corroborating observations include seismic activity, distribution of the Discrete Cosine Transform, quantized JPEG coefficients, and cognition experiments \cite{pietronero,fu2007,gauvrit2009}. To overcome this problem, among the generalizations proposed by several authors, one can highlight the theoretical description introduced by Pietronero \textit{et al.}, mathematically described as below:

\begin{eqnarray}
 P(n) & = & \int_{n}^{n+1} N^{-\alpha} dN \label{eq-pn}
\end{eqnarray}
or by the differential equation:
\begin{eqnarray}
 \frac{dP(N)}{dN} & = & N^{-\alpha}
 \label{eq-pn-diff}
\end{eqnarray}

Solving eq. (\ref{eq-pn-diff}) results in an $\alpha$-logarithm:
\begin{eqnarray}
 P_{\alpha}(n) & = & \frac{1}{1-\alpha}\left[ \left(n+1\right)^{(1-\alpha)} - n^{(1-\alpha)}\right] \\
   & = & n^{(1-\alpha)} \ \ln_{\alpha}\left( \frac{n+1}{n} \right)
 \label{eq-pn-sol}
\end{eqnarray}

According to eq. (\ref{eq-pn-sol}), defined as generalized NBL (gNBL), more frequent first digits than expected by NBL implies $\alpha>1$, while $\alpha<1$ means a first digit frequency below the predicted percentage. As expected, when $\alpha=1$ NBL is fully recovered. Taking $n=d_{1}$ equation (\ref{eq-pn-sol}) is rewritten as:

\begin{eqnarray}
 P_{\alpha}( d_{1} ) & = & d_{1}^{1-\alpha} \ln_{\alpha}\left( \frac{d_{1}+1}{d_{1}}  \right),  \label{gbenford}
\end{eqnarray}

From the approach developed by Pietronero, it is also possible to obtain expressions for the second digit:
\begin{eqnarray}
  P_{\alpha}( d_{2} ) & = & \sum_{d_{1}=1}^{9} (d_{1}d_{2})^{1-\alpha} \ln_{\alpha}\left( \frac{d_{1}d_{2}+1}{d_{1}d_{2}}  \right),    \label{gbenford-2c}
\end{eqnarray}

One could consider gNBL probability for the first-two digits as follows:
\begin{eqnarray}
 P_{\alpha}(d_{1}d_{2}) & = & \displaystyle  \frac{ \left[ \left( d_{1}d_{2}+1 \right)^{1-\alpha}  - \left( d_{1}d_{2} \right)^{1-\alpha}     \right]  }{1-\alpha}, \ \ \ d_{1}d_{2} \in \lbrace 10,11,12,...,99 \rbrace , 
\end{eqnarray}
normalized for each $\alpha$ value.

Table \ref{tab:my-table2} shows each proportional frequency for the first and the second digits.

\begin{table}[]
\caption{Expected proportions for NBL occurrence.}
\label{tab:my-table2}
\resizebox{\textwidth}{!}{%
\begin{tabular}{@{}lccccccccccc@{}}
\toprule
 & Digit & 0 & 1 & 2 & 3 & 4 & 5 & 6 & 7 & 8 & 9 \\ \midrule
\multirow{2}{*}{Position in number} & 1st & - & 0.30103 & 0.17609 & 0.12494 & 0.09691 & 0.07918 & 0.06695 & 0.05799 & 0.05115 & 0.04576 \\
 & 2nd & 0.11968 & 0.11389 & 0.10882 & 0.10433 & 0.10031 & 0.09668 & 0.09337 & 0.09035 & 0.08757 & 0.08500 \\ \bottomrule 
\end{tabular}%
}
\end{table}

\subsection{Assessing conformity to NBL}

An important issue in NBL analysis is to select a convenient statistical method for measuring statistical significance between empirical data and law predicted values. For instance, many investigators traditionally assume both $\chi^{2}$ and $Z$ tests, although both manifest the "excess of power" problem. Thus, to circumvent this effect, Nigrini and Kossovsky suggested the mean absolute deviation (MAD) and sum of squares difference (SSD), respectively \cite{nigrini,kossovsky}. We decided to use both methods in order to enhance the robustness of our statistical analysis, avoiding misinterpretations or superficial conclusions. In this framework, the use of both MAD and SSD enabled Slepkov \textit{et. al} to uncover NBL conformity in answers to every end-of-chapter question in introductory physics and chemistry textbooks \cite{slepkov2015}. Thus, in mathematical form, MAD is defined as:
\begin{eqnarray}
 MAD & = & \frac{ \displaystyle \sum_{i=1}^{n} | AP_{i} - EP_{i}  |}{n}
\end{eqnarray}
where AP is the actual proportion and EP is the expect proportion. In addition, SSD is calculated with the following expression:

\begin{eqnarray}
 SSD & = & \displaystyle \sum_{i=1}^{n} \left( AP_{i} - EP_{i} \right)^{2}\times 10^{4}
\end{eqnarray}

Once again, AP is the actual proportion and EP is the expected proportion. Table \ref{tab.mad1} presents the conformance range for MAD and SSD analysis.

\begin{table}\caption{Ranges of conformity for first, second, and first-two digits.}\label{tab.mad1}
\centering
\resizebox{\textwidth}{!}{%
\begin{tabular}{@{}ccccccc@{}}
\toprule
\multirow{2}{*}{}    & \multicolumn{2}{c}{First Digit}            & \multicolumn{2}{c}{Second Digit}           & \multicolumn{2}{c}{First-two Digits}         \\ \cmidrule(l){2-7} 
                     & Range              & Conclusion            & Range              & Conclusion            & Range                & Conclusion            \\ \midrule
\multirow{4}{*}{MAD} & $0.000$ to $0.006$ & Close conformity      & $0.000$ to $0.008$ & Close conformity      & $0.0000$ to $0.0012$ & Close conformity      \\
                     & $0.006$ to $0.012$ & Acceptable conformity & $0.008$ to $0.010$ & Acceptable conformity & $0.0012$ to $0.0018$ & Acceptable conformity \\
                     & $0.012$ to $0.015$ & Marginal conformity   & $0.010$ to $0.012$ & Marginal conformity   & $0.0018$ to $0.0022$ & Marginal conformity   \\
                     & above $0.015$      & Nonconformity         & above $0.012$      & Nonconformity         & above $0.0022$       & Nonconformity         \\\midrule
\multirow{4}{*}{SSD} & $0$ to $2$         & Close conformity      & $0$ to $2$         & Close conformity      & $0$ to $2$           & Close conformity      \\
                     & $2$ to $25$        & Acceptable conformity & $2$ to $10$        & Acceptable conformity & $2$ to $10$          & Acceptable conformity \\
                     & $25$ to $100$      & Marginal conformity   & $10$ to $50$       & Marginal conformity   & $10$ to $50$         & Marginal conformity   \\
                     & above $100$        & Nonconformity         & above $50$         & Nonconformity         & above $50$           & Nonconformity         \\ \bottomrule
\end{tabular}%
}
\end{table}


\subsection{Distribution fit to MEPP intervals}
Briefly speaking, Weibull and lognormal distributions were fitted to histograms of MEPP intervals applying the method of the maximum likelihood for parameter estimation. The Weibull probability distribution function (pdf) is given by equation (\ref{weibullpdf}):
\begin{eqnarray} \label{weibullpdf}
 f_{wb}(x|r,b) & = & \displaystyle \frac{b}{r}\left(\frac{x}{r}\right)^{b-1}\exp(-(x/r)^{b})
\end{eqnarray}
For $x > 0$, and $r, b \geq 0$. And the lognormal pdf is given by the following:
\begin{eqnarray} \label{lognormalpdf}
 f_{ln}(x|\mu,\sigma) & = & \displaystyle \frac{1}{x \sigma \sqrt{2\pi} } \exp\left(- \frac{(\log(x) - \mu)^{2}}{2\sigma^{2}} \right)
\end{eqnarray}
Where $x, \sigma > 0$.

The likelihood function is the joint probability density, associated with each distribution function, of n identically distributed and independent observations, ${x_{1},...,x_{n}}$, where n is also the sample size. This function gives the probability that a set of observations is best described by a parameter set, ${\theta_{1},...,\theta_{s}}$, from a pdf:

\begin{eqnarray} 
 \mathcal{L}(\theta_{1},...,\theta_{s} |\mathbf{x}) & = & \displaystyle  f(x_{1},...,x_{n}|\theta_{1},...,\theta_{s}) = \prod_{i=1}^{n} f(x_{i}|\theta_{1},...,\theta_{s})
\end{eqnarray}

The maximum likelihood estimator for a parameter $\theta_{i}$ in the parameter set is given by:
\begin{eqnarray} 
\hat{\theta_{i}}(\mathbf{x}) & = & \displaystyle \argmax_{\theta_{i}}  \mathcal{L}(\theta_{1},...,\theta_{s}|\mathbf{x}) 
\end{eqnarray}
or by maximizing the logarithm of the likelihood function (or log-likelihood function)\cite{Greene2017}.

We applied this method to find the parameters from Weibull and lognormal distributions with the highest probability of describing the data. After that, we selected the best model describing the data using, as criteria the Akaike Information Criterion:
\begin{eqnarray} \label{aic}
AIC & = & \displaystyle -2\ln\left(\mathcal{L}(\hat{\theta_{1}},...,\hat{\theta_{s}} |\mathbf{x}\right) +2k
\end{eqnarray}
where k is the number of parameters of the model to be selected, and the Bayesian Information Criterion \cite{Myung2004}:
\begin{eqnarray} \label{bic}
BIC & = & \displaystyle -2\ln\left(\mathcal{L}(\hat{\theta_{1}},...,\hat{\theta_{s}} |\mathbf{x}\right) +k\ln(n)
\end{eqnarray}

\section{Results}
\subsection{Newcomb-Benford law results}

\begin{figure}
\centering
\includegraphics[width=8.0cm]{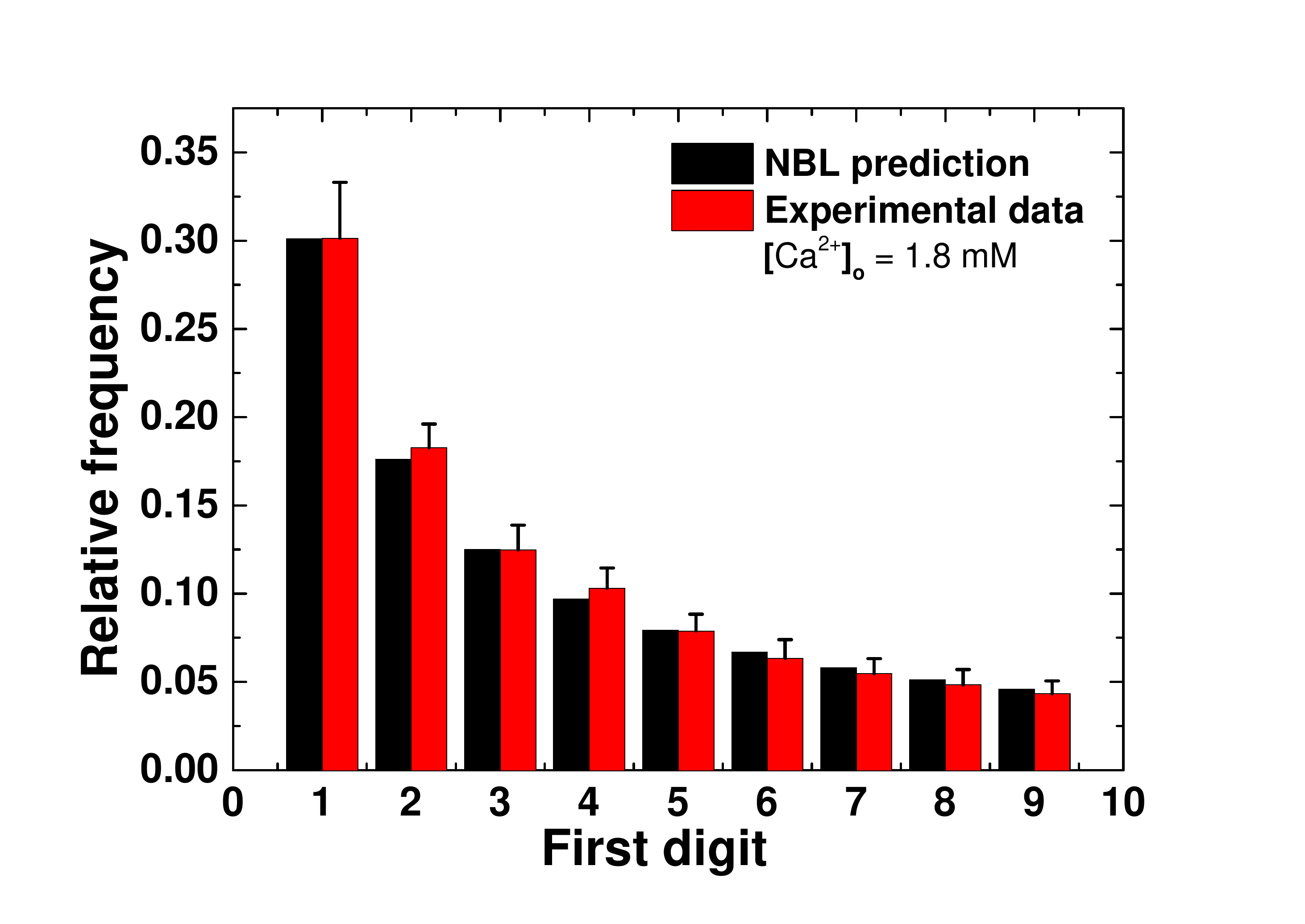}
\includegraphics[width=8.0cm]{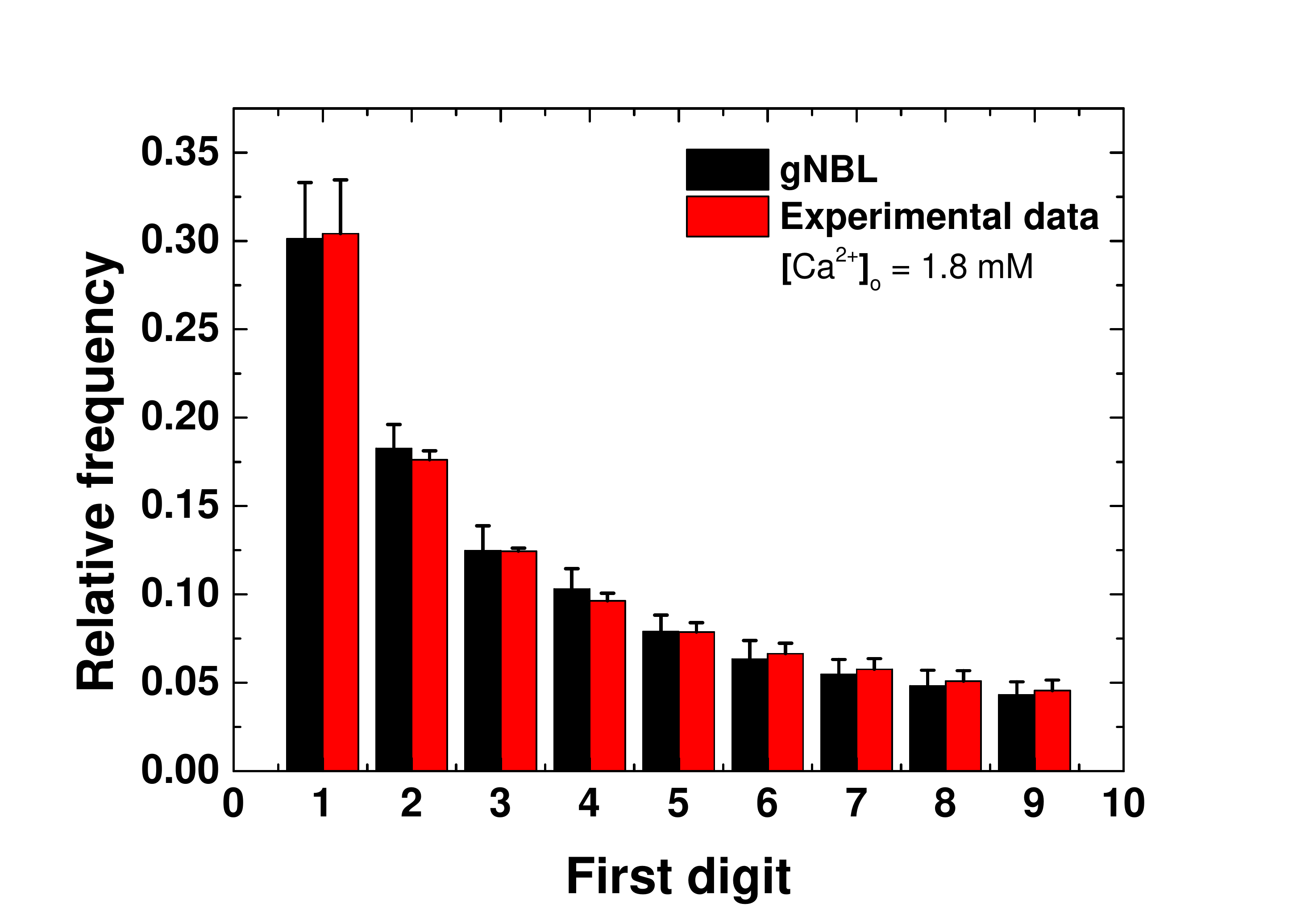}

(a) \hspace{8.0cm} (b)

\includegraphics[width=8cm]{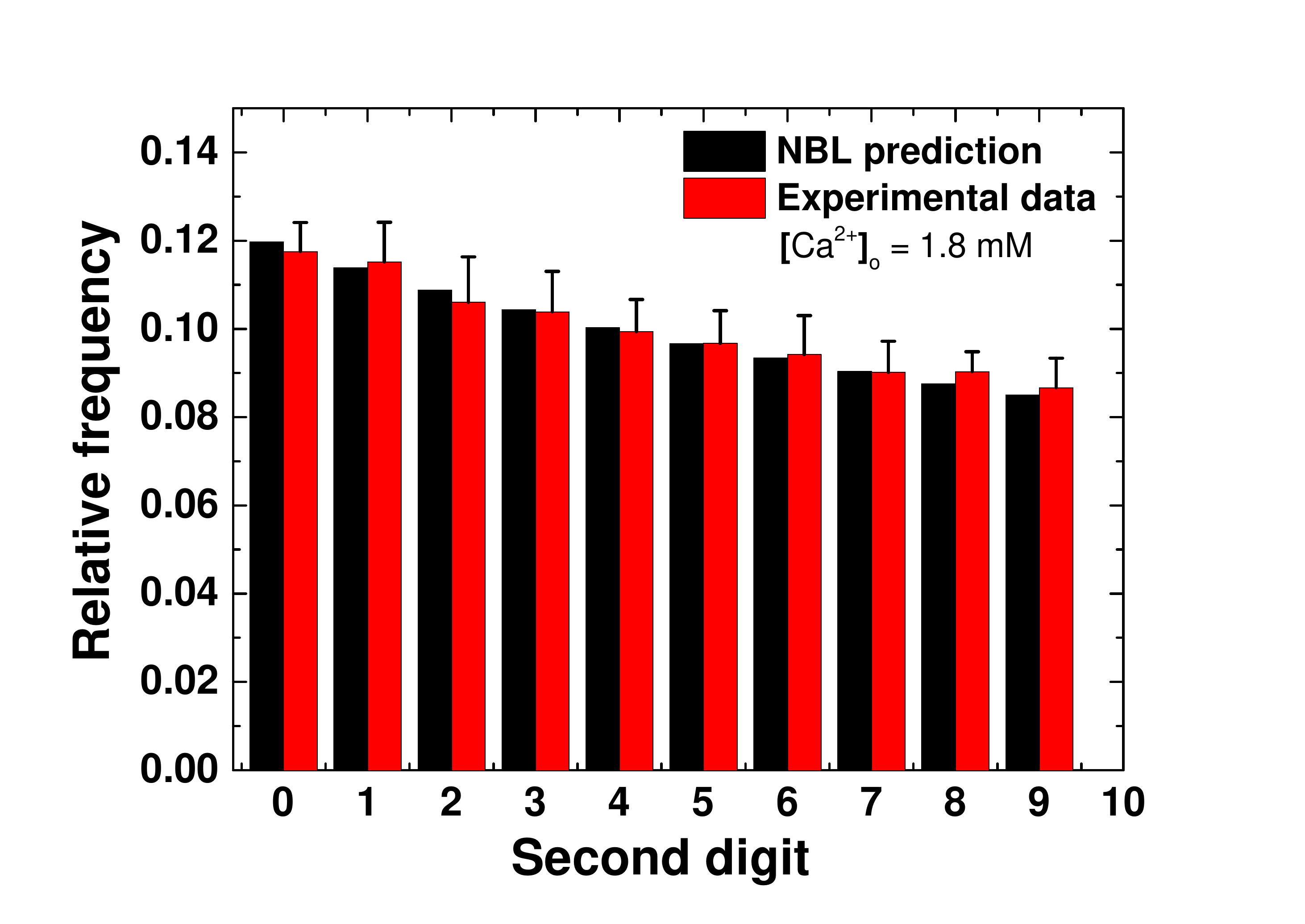}
\includegraphics[width=8cm]{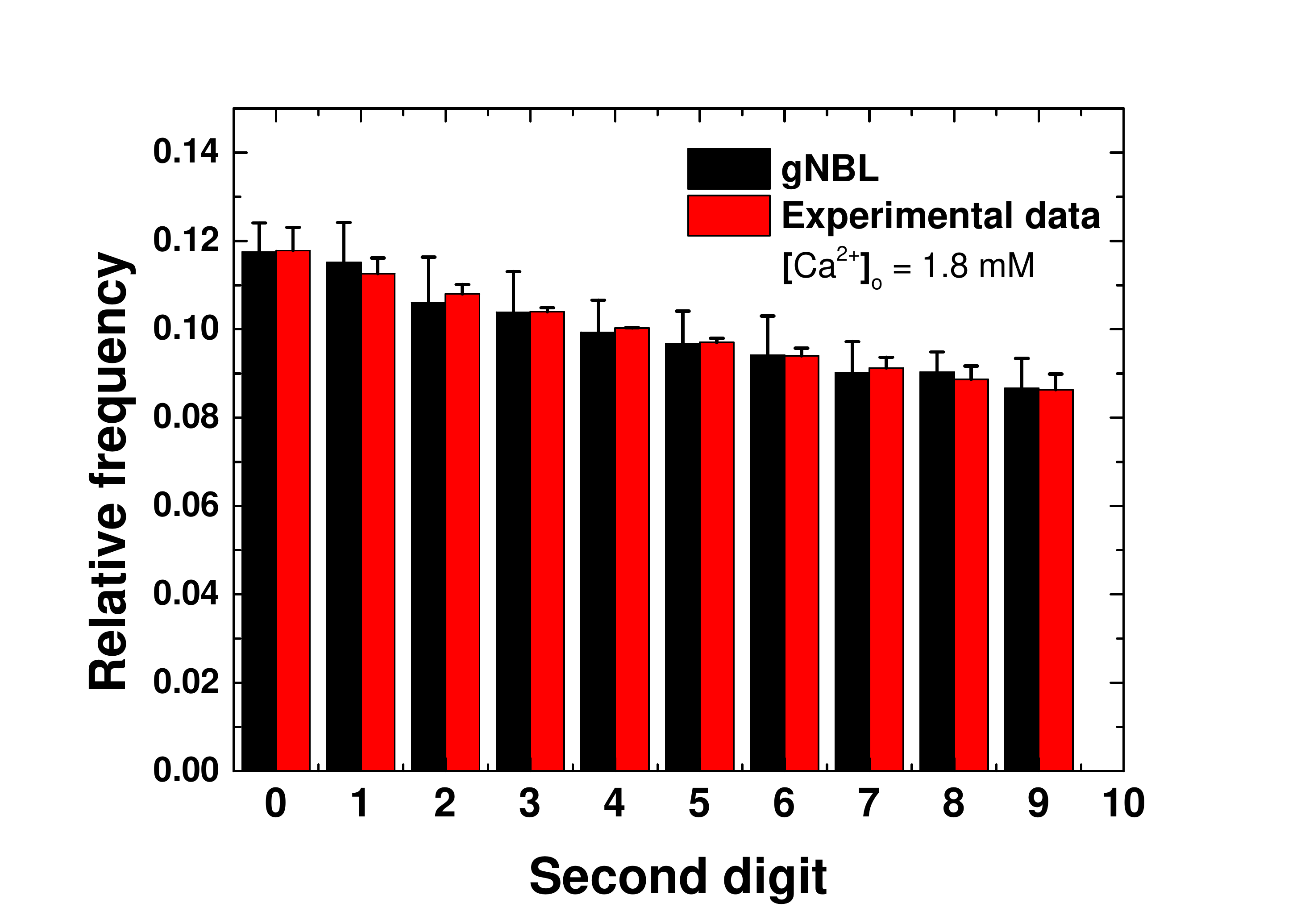}

(c) \hspace{8.0cm} (d)

\includegraphics[width=8cm]{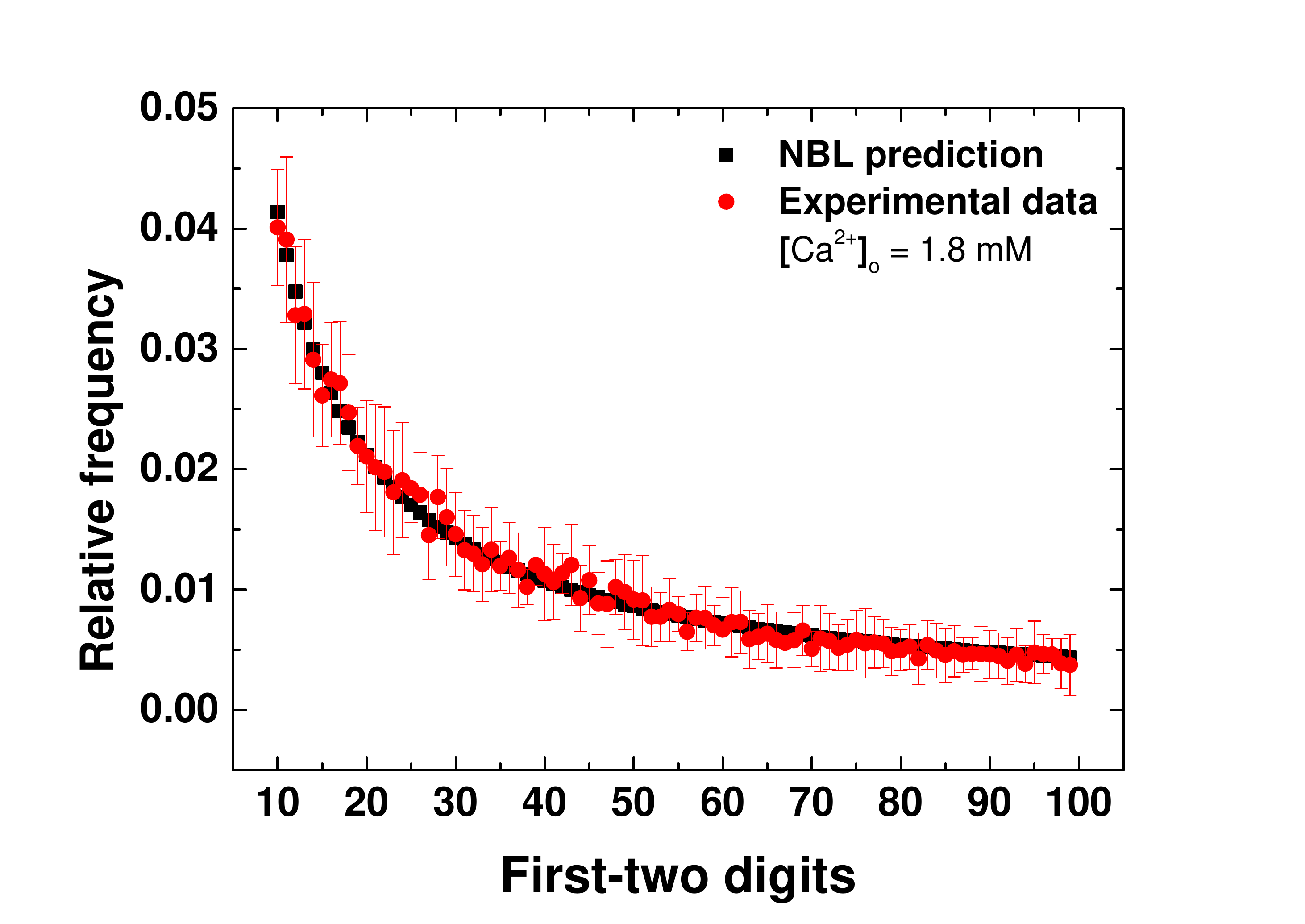}
\includegraphics[width=8cm]{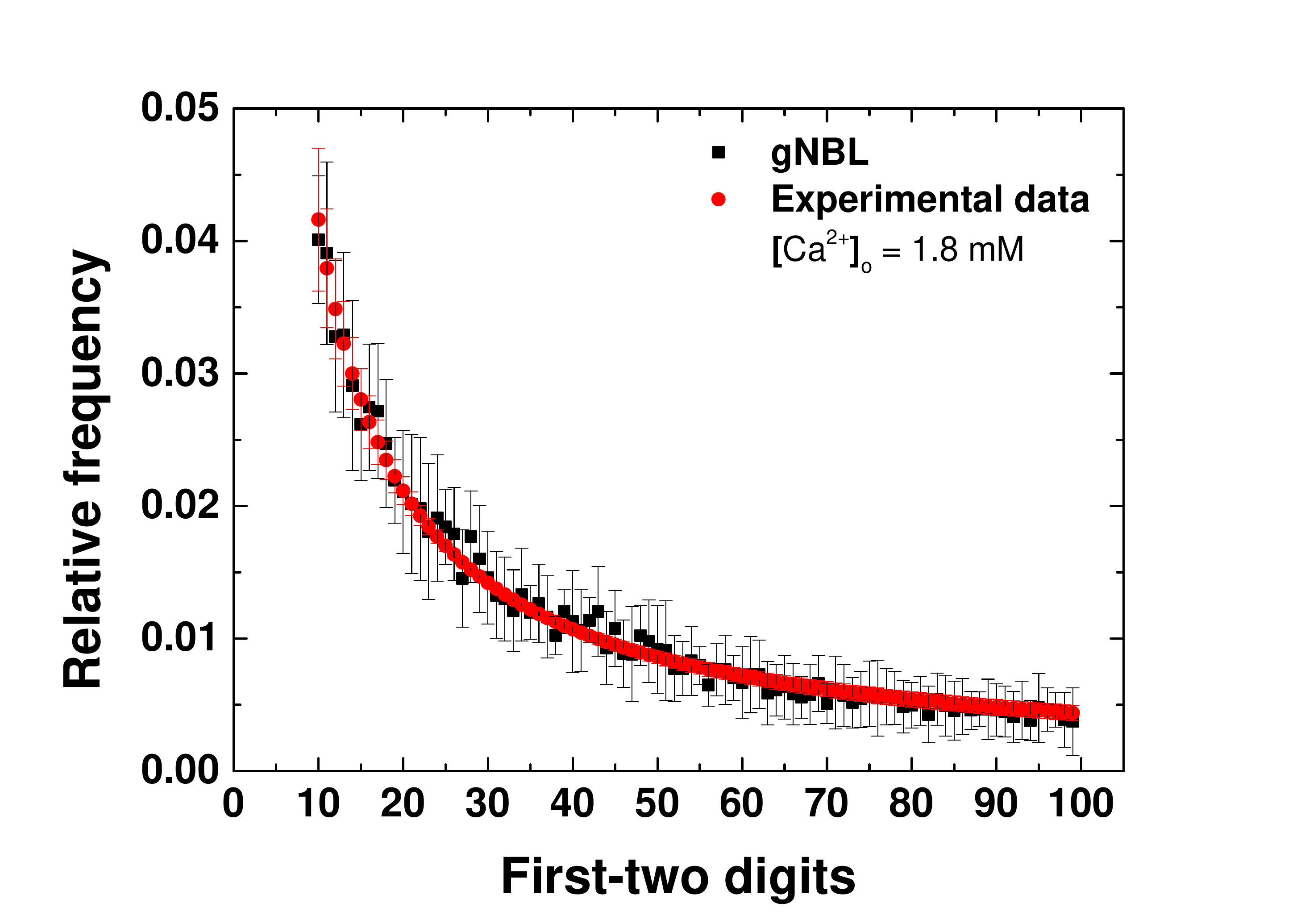}

(e) \hspace{8.0cm} (f)

\caption{Statistical summary (n=14) for physiological $[Ca^{2+}]_{o}$, using NBL and gNBL.} 
\label{fig3}
\end{figure}

The results summarized in tables \ref{tab3} and \ref{tab4}, suggests that the MEPP interval generally follows NBL predictions. We adopted the mean value as a reference value to conclude the existence of conformity. Taling the average value as reference, experimental data studied with MAD and SSD, considering NBL for first and second digit, provided ambiguous results given by different conformance degrees. When assuming MAD, for first-two digits results, both NBL e gNBL pointed to non-conformities, while in analysis performed with SSD all data achieved at least a marginal conformity. Figure \ref{fig3} presents a summary for statistical analysis, considering $[Ca^{2+}]_{o}$ at physiological levels, showing both NBL e gNBL statistical descriptions. A visual inspection enables to observe an excellent agreement of experimental data and predicted values. According to our calculations, taking the first digit, gNBL gives a slightly better conformity for $[Ca^{2+}]_{o}< 1.8$ mM, while above this concentration (with exception of $[Ca^{2+}]_{o}= 15$ mM), both laws seem to perform equivalently. Moreover, NBL and gNBL are equivalent for second digit analysis, with exception of $[Ca^{2+}]_{o}= 2.4$ mM, which SSD pointed a better gNBL corformity. In summary, these findings show that MEPP time series follow the first and second digit phenomenon. On the other hand, experimental data followed the first-two digits proportions only when SSD was assumed or for $[Ca^{2+}]_{o} \ge 10$ mM. These results suggest that MEPP time series does not perfectly conforms NBL or gNBL for first-two digits levels. 

\begin{table}[]
\caption{Statistical resume of conformity analysis, where the conformance level is concluded taking the mean value as reference.}\label{tab3} 
\begin{adjustbox}{max width=\textwidth}
\begin{tabular}{@{}ccccccccl@{}}
\toprule
\multicolumn{1}{c}{\multirow{4}{*}{\textbf{\begin{tabular}[c]{@{}c@{}}Concentration\\  (mM)\end{tabular}}}} & \multicolumn{8}{c}{\textbf{First Digit}}                                                                                                                                                                                                                      \\ \cmidrule(l){2-9} 
\multicolumn{1}{c}{} & \multicolumn{2}{c}{\textbf{NBL - MAD}}  & \multicolumn{2}{c}{\textbf{gNBL -  MAD}}  
                     & \multicolumn{2}{c}{\textbf{NBL - SSD}}  & \multicolumn{2}{c}{\textbf{ gNBL - SSD}} \\ 
\cmidrule(l){2-9} 
\multicolumn{1}{c}{} & \multicolumn{1}{c}{Value} & \multicolumn{1}{c}{Conformity} & \multicolumn{1}{c}{Value} & \multicolumn{1}{c}{Conformity} & \multicolumn{1}{c}{Value} & \multicolumn{1}{c}{Conformity} & \multicolumn{1}{c}{Value} & \multicolumn{1}{c}{Conformity} \\ \midrule
0.6                                       
& $0.01206\pm0.00481$& Marginal & $0.01048\pm0.00443$ & Acceptable  & $25.67505\pm21.01661$ & Marginal & $15.43984\pm14.01542$ & Acceptable  \\
1.2                                        
& $0.01207\pm0.00276$& Marginal & $0.01030\pm0.00330$ & Acceptable  & $21.18447\pm12.32537$ & Acceptable & $16.02790\pm12.41851$ & Acceptable  \\
1.8                                       
& $0.01054\pm0.00324$& Acceptable & $0.00773\pm0.00229$ & Acceptable & $19.07781\pm20.20499$ & Acceptable & $8.36754\pm4.27991$ & Acceptable  \\
2.4                                       
& $0.00989\pm0.00167$& Acceptable & $0.00949\pm0.00174$ & Acceptable & $15.20449\pm7.00340$ & Acceptable  & $12.16417\pm4.17319$ & Acceptable  \\
4.8                                       
& $0.01112\pm0.00426$& Acceptable & $0.00834\pm0.00288$ & Acceptable & $22.75867\pm23.40724$ & Acceptable  & $9.74163\pm7.42517$ & Acceptable  \\
10                                       
& $0.00885\pm0.00167$& Acceptable & $0.00764\pm0.00200$   & Acceptable & $11.20839\pm3.75458$ & Acceptable  & $7.86167\pm3.43018$ & Acceptable  \\
15                                        
& $0.00778\pm0.00188$& Acceptable & $0.00583\pm0.00175$ & Close  & $9.25291\pm5.18910$& Acceptable  & $4.77294\pm2.67565$ & Acceptable  
\\   
\bottomrule

\toprule
\multirow{3}{*}{\begin{tabular}[c]{@{}c@{}}Concentration \\ (mM)\end{tabular}} & \multicolumn{8}{c}{Second Digit}                                                                                                                                                                          \\ \cmidrule(l){2-9} 
& \multicolumn{2}{c}{NBL -  MAD}  & \multicolumn{2}{c}{gNBL - MAD}  & \multicolumn{2}{c}{NBL - SSD}  & \multicolumn{2}{c}{gNBL - SSD}                   \\ \cmidrule(l){2-9} 
& Value & Conformance     & Value & Conformance    & Value & Conformance      & Value & Conformance \\ 
\midrule
0.6                                                                            
& $0.00899\pm0.00335$ & Acceptable & $0.00823\pm0.00326$ & Acceptable & $13.85888\pm10.55529$ & Marginal & $11.50431\pm8.61014$  & Marginal    \\
1.2                                                                            
& $0.00733\pm0.00577$  & Close     & $0.00675\pm0.00513$ & Close      & $12.22850\pm18.44405$  & Marginal    & $10.15934\pm15.22852$ & Marginal    \\
1.8                                                                            
& $0.00609\pm0.00168$     & Close     & $0.00577\pm0.00172$ & Close    & $6.00561\pm3.31274$   & Acceptable  & $5.18699\pm2.92581$   & Acceptable  \\
2.4                                                                            
& $0.00886\pm0.00152$     & Acceptable  & $0.00806\pm0.00162$ & Acceptable & $11.41090\pm2.73660$    & Marginal    & $9.88293\pm2.89251$   & Acceptable  \\
4.8                                                                            
& $0.00737\pm0.00082$ & Close       & $0.00743\pm0.00114$ & Close      & $8.64290\pm1.77500$      & Acceptable  & $8.24204\pm2.12804$   & Acceptable  \\
10                                                                            
& $0.00574\pm0.00143$     & Close       & $0.00550\pm0.00148$  & Close      & $5.28398\pm2.39307$   & Acceptable  & $4.76852\pm2.38107$   & Acceptable  \\
15                                                                             
& $0.00562\pm0.00204$     & Close       & $0.00512\pm0.00213$ & Close      & $4.74227\pm3.35831$   & Acceptable  & $4.06391\pm3.50945$   & Acceptable  \\ \bottomrule

\toprule
\multirow{3}{*}{\begin{tabular}[c]{@{}c@{}}Concentration \\ (mM)\end{tabular}} & \multicolumn{8}{c}{First-two Digits}                                                                                                                                                                          \\ \cmidrule(l){2-9} 
& \multicolumn{2}{c}{NBL -  MAD}  & \multicolumn{2}{c}{gNBL - MAD}  & \multicolumn{2}{c}{NBL - SSD}  & \multicolumn{2}{c}{gNBL - SSD}                   \\ \cmidrule(l){2-9} 
& Value & Conformance     & Value & Conformance    & Value & Conformance      & Value & Conformance \\ 
\midrule
0.6                                                                         & $0.00314\pm0.00113$     & Nonconformity & $0.00309\pm0.00107$     & Nonconformity & $18.06113\pm13.42578$ & Marginal  & $16.89847\pm12.39917$ & Marginal    \\
1.2                                                                         & $0.00277\pm0.00140$      & Nonconformity & $0.00273\pm0.00141$     & Nonconformity & $14.26131\pm16.62761$ & Marginal  & $13.83585\pm16.66196$ & Marginal    \\
1.8                                                                         & $0.00232\pm0.00058$ & Nonconformity & $0.00221\pm0.00053$ & Nonconformity      & $8.88829\pm4.54648$   & Acceptable  & $7.77473\pm3.55850$    & Acceptable  \\
2.4                                                                        & $0.00300\pm0.00041$   & Nonconformity & $0.00297\pm0.00041$  & Nonconformity & $13.79984\pm3.84245$  & Marginal    & $13.27410\pm3.74433$   & Marginal    \\
4.8                                                                         & $0.00253\pm0.00043$  & Nonconformity & $0.00239\pm0.00023$ & Nonconformity & $10.71470\pm3.61727$  & Marginal    & $8.85734\pm1.49726$   & Acceptable  \\
10                                                                          & $0.00180\pm0.00029$  & Acceptable    & $0.00179\pm0.00027$ & Acceptable    & $5.52777\pm2.02036$   & Acceptable  & $5.26379\pm1.87770$    & Acceptable  \\
15                                                                          & $0.00193\pm0.00063$ & Marginal      & $0.00187\pm0.00061$ & Marginal      & $6.19024\pm4.45760$    & Acceptable  & $5.78632\pm4.23051$   & Acceptable  \\ \bottomrule
\end{tabular}
\end{adjustbox}
\end{table}

The assumption of gNBL gave $\alpha$ values associated to each concentration. On average, for $[Ca^{2+}]_{o}=  0.6$ mM we obtained $\alpha < 1$, which can be attributed to smaller time series, since during low $[Ca^{2+}]_{o}$ administration, the MEPP firing is attenuated. In contrast, $[Ca^{2+}]_{o}= 1.8$ mM provided $\alpha = 1$, showing that considering the standard deviation, at least in physiological concentrations, NBL analysis is fully accomplished. Furthermore, data  for $[Ca^{2+}]_{o}=(2.4-4.8)$ mM resulted $\alpha \neq 1$ for the three digits analysis, showing that gNBL is more appropriate in both concentrations. Finally, $[Ca^{2+}]_{o} = (10-15)$ mM gave $\alpha \cong 1$, again demonstrating that NBL is suitable to investigate MEPP discharge at higher frequencies. 

\begin{table}[]
\caption{Statistical analysis for $\alpha$ parameter for each $[Ca^{2+}]_{o}$.}\label{tab4} 
\begin{tabular}{@{}cccc@{}}
\toprule
$[Ca^{2+}]_{o}$ (mM) & First Digit                      & Second Digit                     & First-two Digits                 \\ \midrule
0.6                & $0.93095\pm0.09851$ & $0.84033\pm0.31991$ & $0.91575\pm0.08363$ \\
1.2                & $0.98249\pm0.08515$ & $1.04133\pm0.28682$ & $0.97653\pm0.06864$ \\
1.8                & $1.01088\pm0.12400$ & $0.92119\pm0.19497$ & $1.00045\pm0.11613$ \\
2.4                & $0.98163\pm0.06584$ & $0.82991\pm0.25890$ & $0.97604\pm0.07917$ \\
4.8                & $1.11593\pm0.07896$ & $1.11977\pm0.16492$ & $1.11914\pm0.08594$ \\
10                 & $1.03481\pm0.06022$ & $1.02929\pm0.14418$ & $1.03750\pm0.04183$  \\
15                 & $1.02135\pm0.07858$ & $1.03680\pm0.16571$ & $1.02319\pm0.06660$  \\ \bottomrule
\end{tabular}
\end{table}

\subsection{Distribution fitting results}
We analysed a set of 78 electrophysiological recordings, verifying which distribution, Weibull or lognormal, best fitted those data. Applying both Akaike and Bayesian information criterion, we find that 76 recordings were best fitted by the Weibull distribution. The only exceptions were two recordings collected at $[Ca^{2+}]_{o}=1.2$ mM and $[Ca^{2+}]_{o}=15$ mM, respectively. A table with statistical data from fittings is presented in the supplementary material. Figure \ref{fig4} shows interval histograms fitted by Weibull and lognormal distributions for three different $[Ca^{2+}]_{o}$.

\begin{figure}
\centering
\includegraphics[width=8cm]{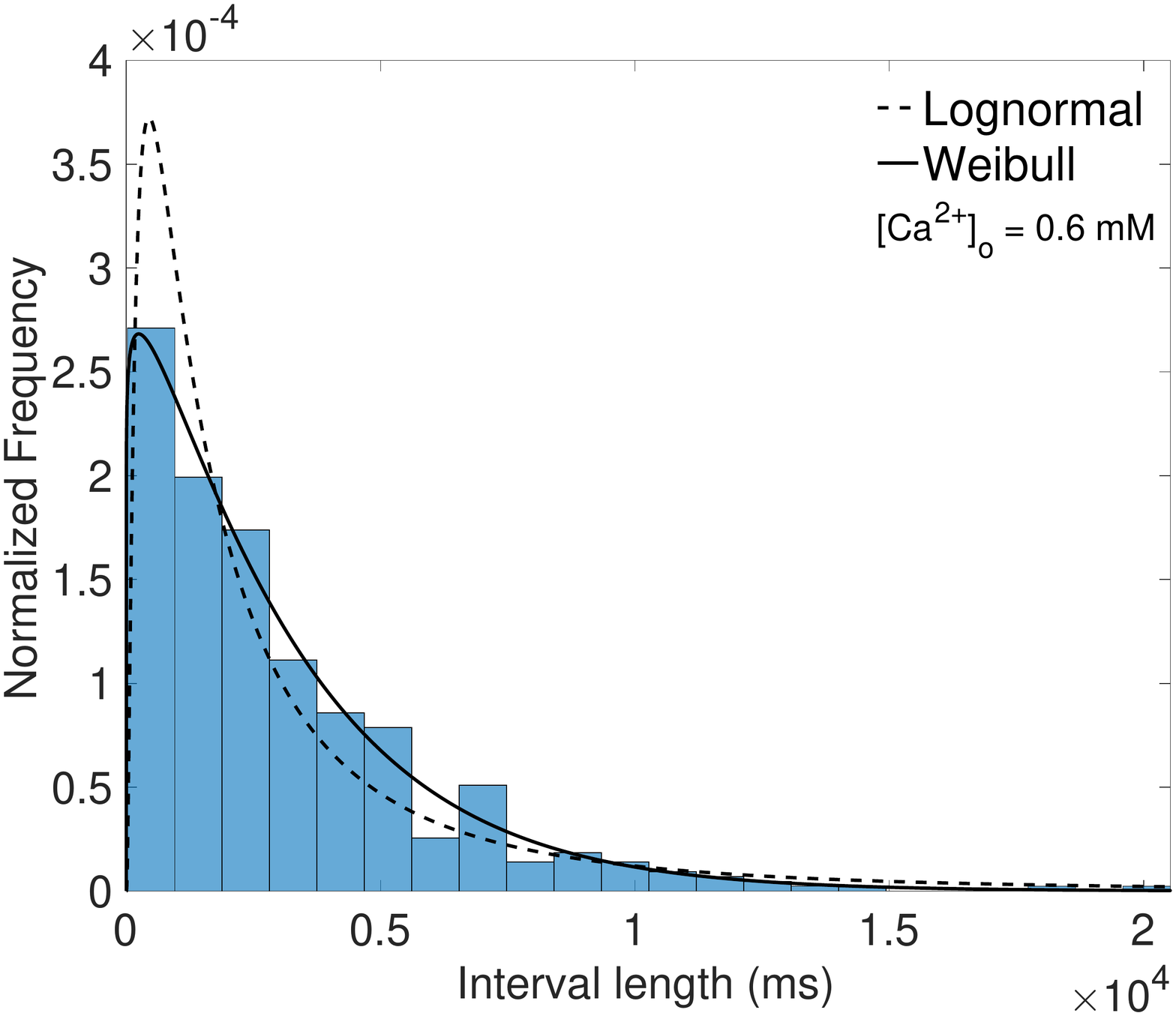}
\includegraphics[width=8cm]{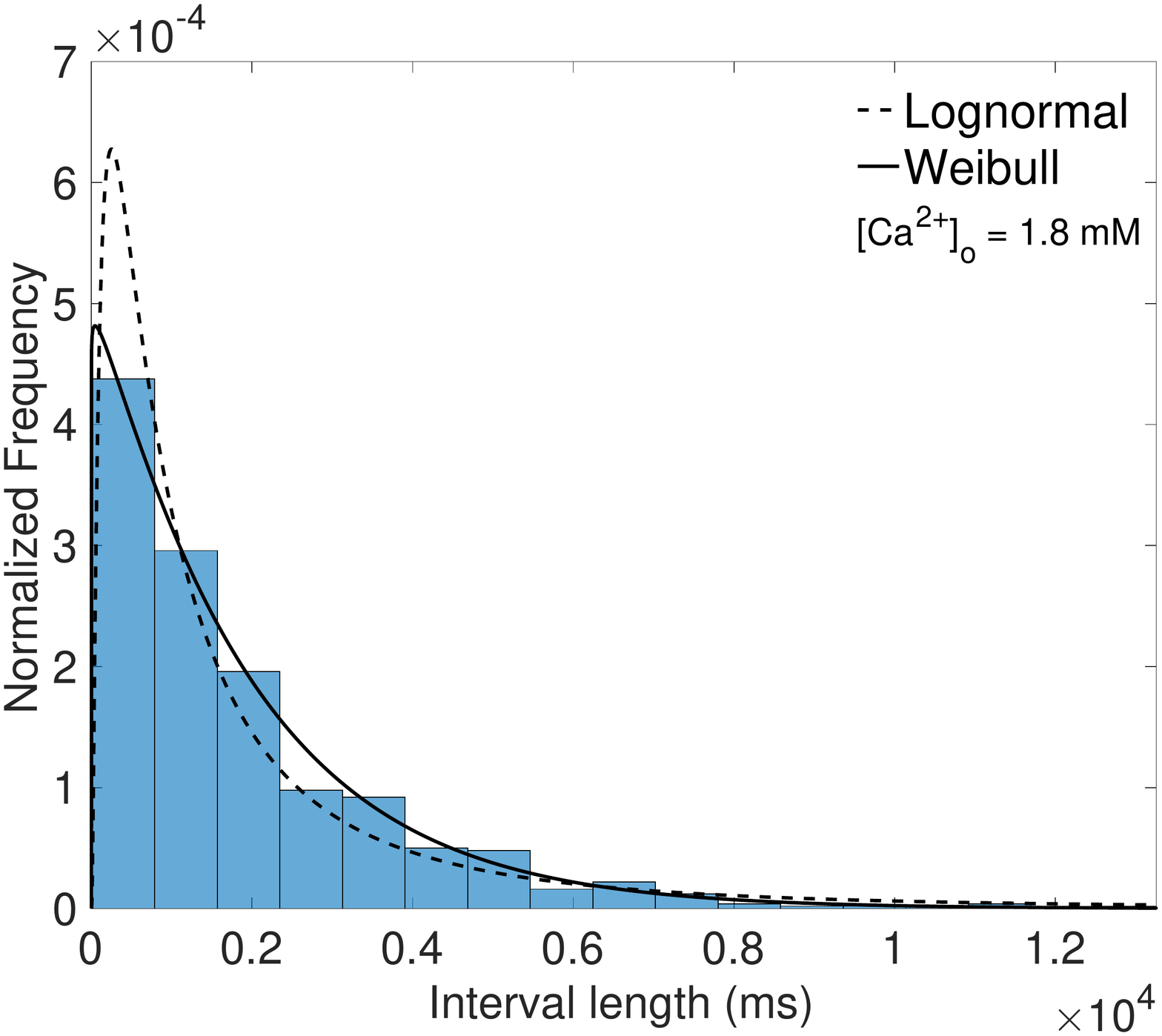}
(a) \hspace{8.0cm} (b)
\includegraphics[width=8cm]{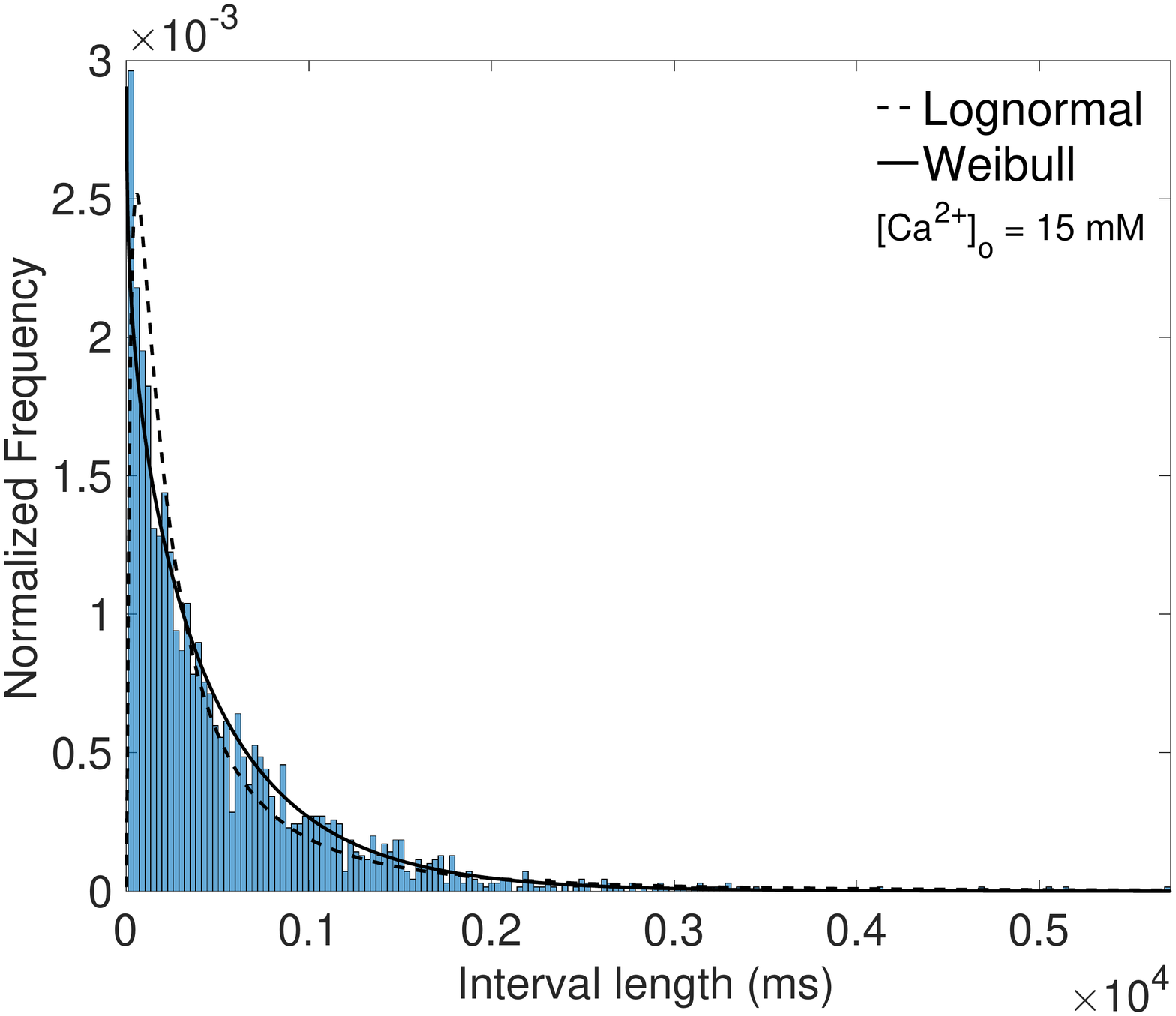}
 \hspace{8.0cm} (c)
\caption{Weibull and lognormal fits for MEPP interval distributions for three representative $[Ca^{2+}]_{o}$. The adjusted parameters from the best fitted distributions (Weibull) are: r = 3125.813 and b = 1.070 (0.6 mM), r = 1910.256 and b = 1.022 (1.8 mM), r = 506.781 and b = 0.925 (15 mM).}  
\label{fig4}
\end{figure}

\section{Discussion}

In the present work, successful application of NBL and gNBL strengthen previous evidences showing that MEPP dynamics is governed by scale invariance laws. To prudently tackle this problem, $[Ca^{2+}]_{o}$ was adjusted in seven concentrations, where conformity with first, second, and first-two digit pattern was investigated. Our results strongly suggest that spontaneous secretion of neurotransmitters, in the mammalian diaphragm, obeys Benford's law predictions in a very satisfactory manner. It is important to highlight that a visual inspection of all experimental data already showed the typical asymmetric shape, characterized by a long tail and heavy skewness toward the smaller digits. Despite this common feature shared by all data, curiously certain recordings did not reach statistical conformity. This raised a question about what physiological reason could be associated with these unexpected findings. It is well known that membrane potential fluctuations impose changes in MEPP rate. In fact, membrane depolarizations increase, while hyperpolarization decreases the frequency of these events. Moreover, there is a $Ca^{2+}$ oscillation within the synaptic terminal, which is modulated by the extracellular content of this ion \cite{Rahamimoff1978}. Such oscillations will reflect in a periodical fluctuation of MEPP discharge as well, which could not only favours short intervals or, in NBL framework, smaller digits. In particular, the fluctuations described above also may be manifested in $\alpha \neq 1$, attested by deviations observed from the classical NBL prediction. Although we indicate probable biological scenarios for observing nonconformity and deviations from the classical NBL in certain data, further studies are required to investigate this interpretation.  

Regulatory mechanisms for neurotransmitter releasing require a complex network of molecular cascades, being regulated by extracellular ions. Exocytosis rate may be exacerbated by increasing $[Ca^{2+}]_{o}$ in the Ringer solution. According to our \textit{in vitro} research, we did not observe a relation between $[Ca^{2+}]_{o}$ and conformity level. From these observations, one can formulate: could \textit{in vitro} results be extrapolated to \textit{in vivo} conditions? Indeed, an important debate in electrophysiology concerns to question if electrical activity extracted from \textit{in vitro} recordings corresponds to the \textit{in vivo} conditions. Among the disadvantages we find that \textit{in vitro} produce mechanical stress in tissues, introducing electrical artefacts and biochemical changes. On the other hand, this category also has many advantages such as the isolation of the tissue or cell, allowing easy local pharmacological manipulation, elimination of afferent contribution from other body areas, among others. For these reasons, isolated diaphragm tissue is also amenable to NBL verification when pH and temperature are locally modified. Many reports documented that electrical activity of synaptic terminals is dependent on thermal and acidification level \cite{muniak1982}. Both parameters also govern MEPP frequency as well. An important physiological consequence of temperature elevation is the acceleration of the vesicular fusion rate, reflected in the MEPP frequency increment. In the future, we expect to expand this study in order to address how both physiological parameters may interfere in the NBL conformance. 

The gradual increment in $[Ca^{2+}]_{o}$ evoked the expected elevation of MEPP frequency, allowing us to verify the validity of NBL/gNBL for different sample size intervals or physiological conditions. This procedure uncovered cases where the statistical analysis gave different levels of conformity or did not even indicate conformity. Thus, our work revealed a rich statistical  scenario. We had already pointed to a physiological substrate to explain the conformance heterogeneity. In mathematical terms, a complementary answer may be achieved reasoning around gNBL. In general, independently of the digit, using gNBL gave  more frequent conformity than NBL. Due to this heterogeneous scenario, one can advocate that the best strategy for studying MEPP time series consists in adopting both NBL and gNBL in conjunction with MAD and SSD tests. We also recommend to consider performing a first, second, and first two-digits calculations in order to achieve a more careful conclusion. 

A relationship is reported between NBL and long range correlation phenomenon described in terms of the nonextensive theory proposed by Tsallis \cite{shao}. In this framework, Shao and Ma carried out a theoretical study associating NBL in the nonextensive context. These authors stated that  NBL confirmation in different systems is theoretically expected, at least, in systems obeying nonextensive statistics. Relative to NMJ, Silva \textit{et al.} showed that MEPP histograms are better understood when adjusted with long tail functions, for instance, a q-Gaussian distribution \cite{adjesbr}. Still according to this study, application of Detrended Fluctuation Analysis (DFA) also strengthened the scenario for the existence of long-range correlations in MEPP intervals. Since DFA allows detection of scale-invariance embedded in time series data, the present results merge with these previous reports. Yet, the relationship between NBL is well studied with other distribution functions such as lognormal and Weibull distributions \cite{formam}. Also motivated by their utility in neurophysiology, we decide to perform simulations assuming both functions. Our experimental preparation and theoretical design confirmed a theoretical study showing that Weibull distribution is the more suitable function than the lognormal distribution.The findings also represent an important empirical support in favour of a close relation between NBL and the Weibull function. Consequently, another important result, is that the Weibull distribution characterized the best statistical modelling for describing MEPP time series. Li offered an intrinsic plasticity model, demonstrating that the probability distribution of the neuronal firing is better explained using the Weibull distribution \cite{Li2011}. Along with this results, Weibull function also appeared as very useful statistics for investigating the neuronal firing of primary olfactory system and single \textit{locus coeruleus} neurons \cite{Park2014,Camproux1996}. Our results reinforce the importance of Weibull statistics in the quantitative analysis in context of communication between nerve and muscle. Thus, a next step concerns to employ Weibull statistics to quantify the plasticity mechanisms at the diaphragm NMJ.  

Finally, Bormashenko asked why NBL is frequently observed in statistical data \cite{Bormashenko2016}. According to his view, like NBL, many systems entropically governed are described by a logarithm dependence. This argument may explain why the first digit phenomenon is so frequently observed in different systems and conditions, including the results here described. It is worth mentioning that previous work, carried out in amphibian NMJ, also reported the intervals distribution described by a logarithmic behaviour \cite{takeda}. Another intriguing question was also formulated by Lemons and Kossovsky, which asked why there are more small things in the world than large things \cite{lemons1985,kossovsky}. In keeping with these authors one can paraphrase: Why does short MEPP interval, given by the abundance of first digit, prevail among the other ones? In the NMJ diaphragm there are thousands of crowded vesicles, with a diameter of 50 nanometers, sharing a confined space. When combined, high density of vesicles and reduced spatial dimension, certainly favours a higher likelihood for vesicle fusion, corroborating for short MEPP intervals existence. Thus, NBL/gNBL descriptions represent an elegant methodology for assessing interesting features of the probabilistic nature of neurotransmission \cite{branco}. Furthermore, shorter intervals still represent a cellular mechanism to avoid receptor recruitment from the cellular membrane. Indeed, Saitoe \textit{et al.} showed that absence of the glutamate receptor in \textit{Drosophila} is directly related to lacking spontaneous transmitter release. Strengthening this reflection, McKinney \textit{et al.} also demonstrated that miniature synaptic events are required in order to maintain dendritic spines in hippocampal slice. Prompted by these findings, one can hypothesize that base or scale invariance are important requisites in preventing the synapse collapse. Finally, the present results also reinforce that MEPP activity organizes itself in time as scale-invariant phenomenon. However, how exocytotic machinery can be orchestrated into a base/scale-invariant behaviour will be the focus of further research. 

\section{Concluding remarks}

To the best of our knowledge, this is the first work suggesting that spontaneous synaptic transmission obeys NBL. In addition, MEPP intervals showed conformity with the NBL independtly of $[Ca^{2+}]_{o}$. In this context, NBL remained valid in hypercalcemia and hypocalcemia conditions. We next showed that, when compared to lognormal distribution, Weibull statistics is more appropriate to adjust MEPP intervals from diaphragm electrophysiological recordings. We hope to extend the present research including both excitatory and inhibitory synapses in the brain. Moreover, it could also be relevant to examine a possible existence of the anomalous number phenomenon from NMJ of non-mammalian species and pathological tissues, such as during administration of toxins and drugs. 

\section{Acknowledgements}

We are in debt to Alex Ely Kossovsky for his discussion and valuable help during the preparation of this work. The authors would like to thank the Multi-User Facility of Drug Research and Development Center of Federal University of Cear\'a for technical support. We also to thank Dr. Barbara Piechocinska for reading the manuscript.

\printcredits

\bibliographystyle{unsrt}

\bibliography{ManuscriptBenford070619}

\begin{thebibliography}{10}

\bibitem{sanes}
J.R. Sanes and J.W. Lichtman.
\newblock Development of the vertebrate neuromuscular junction.
\newblock {\em Annual Review of Neuroscience}, 22(1):389--442, mar 1999.

\bibitem{katz1}
B.~Katz.
\newblock Neural transmitter release: from quantal secretion to exocytosis and
  beyond.
\newblock {\em Journal of Neurocytology}, 25(1):677--686, jan 1996.

\bibitem{katz2}
P.~Fatt and B.~Katz.
\newblock Spontaneous subthreshold activity at motor nerve endings.
\newblock {\em J. Physiol.}, 117(1):109--128, 1952.

\bibitem{bennett}
M.R. Bennett and J.L. Kearns.
\newblock Statistics of transmitter release at nerve terminals.
\newblock {\em Progress in Neurobiology}, 60(6):545--606, apr 2000.

\bibitem{Higashima1990}
M.~Higashima, S.~Sawada, and C.~Yamamoto.
\newblock Applicability of pascal distribution to quantal analysis for
  non-stationary release of neurotransmitter.
\newblock {\em Neuroscience Letters}, 115(2-3):231--236, jul 1990.

\bibitem{washio}
H.M. Washio and S.T. Inouye.
\newblock The statistical analysis of spontaneous transmitter release at
  individual junctions on cockroach muscle.
\newblock {\em J. Exp. Biol.}, 87:195--201, 1980.

\bibitem{lowen}
S.B. Lowen, S.S. Cash, M.M., and M.C. Teich.
\newblock Quantal neurotransmitter secretion rate exhibits fractal behavior.
\newblock {\em The Journal of Neuroscience}, 17(15):5666--5677, aug 1997.

\bibitem{takeda}
T.~Takeda, A.~Sakata, and M.~Takahide.
\newblock Fractal dimensions in the occurrence of miniature end-plate potential
  in a vertebrate neuromuscular junction.
\newblock {\em Progress in Neuro-Psychopharmacology and Biological Psychiatry},
  23(6):1157--1169, aug 1999.

\bibitem{adjesbr}
A.~J. da~Silva, R.~F. Lima, and M.~A. Moret.
\newblock Nonextensivity and self-affinity in the mammalian neuromuscular
  junction.
\newblock {\em Physical Review E}, 84(4), oct 2011.

\bibitem{kloot1989}
W.~Van der Kloot.
\newblock Statistical and graphical methods for testing the hypothesis that
  quanta are made up of subunits.
\newblock {\em Journal of Neuroscience Methods}, 27(1):81--89, feb 1989.

\bibitem{robinson1976}
J.~Robinson.
\newblock Estimation of parameters for a model of transmitter release at
  synapses.
\newblock {\em Biometrics}, 32(1):61, mar 1976.

\bibitem{Kamiya1992}
H.Kamiya, S.Sawada, and C.Yamamoto.
\newblock An empirical test for the reliability of quantal analysis based on
  pascal statistics.
\newblock {\em Journal of Neuroscience Methods}, 42(1-2):19--26, apr 1992.

\bibitem{McKeegan2002}
D.E.F. McKeegan.
\newblock Spontaneous and odour evoked activity in single avian olfactory bulb
  neurones.
\newblock {\em Brain Research}, 929(1):48--58, mar 2002.

\bibitem{buzsaki}
G.~Buzs{\'{a}}ki and K.~Mizuseki.
\newblock The log-dynamic brain: how skewed distributions affect network
  operations.
\newblock {\em Nature Reviews Neuroscience}, 15(4):264--278, feb 2014.

\bibitem{Levine1991}
M.~W. Levine.
\newblock The distribution of the intervals between neural impulses in the
  maintained discharges of retinal ganglion cells.
\newblock {\em Biological Cybernetics}, 65(6):459--467, oct 1991.

\bibitem{rusakov2006}
D.A. Rusakov.
\newblock Ca2+-dependent mechanisms of presynaptic control at central synapses.
\newblock {\em The Neuroscientist}, 12(4):317--326, aug 2006.

\bibitem{newcomb}
S.~Newcomb.
\newblock Note on the frequency of use of the different digits in natural
  numbers.
\newblock {\em American Journal of Mathematics}, 4(1/4):39, 1881.

\bibitem{benford}
F.~Benford.
\newblock The law of anomalous numbers.
\newblock {\em Proc. Am. Philos. Soc.}, 78(4):551--572, 1938.

\bibitem{Diekmann2007}
A.~Diekmann.
\newblock Not the first digit! using benford{\textquotesingle}s law to detect
  fraudulent scientif ic data.
\newblock {\em Journal of Applied Statistics}, 34(3):321--329, apr 2007.

\bibitem{Miller2009}
M.J. Nigrini and S.J. Miller.
\newblock Data diagnostics using second-order tests of
  benford{\textquotesingle}s law.
\newblock {\em Auditing: A Journal of Practice {\&} Theory}, 28(2):305--324,
  nov 2009.

\bibitem{miller2007}
M.J. Nigrini and S.J. Miller.
\newblock Benford's law applied to hydrology data{\textemdash}results and
  relevance to other geophysical data.
\newblock {\em Mathematical Geology}, 39(5):469--490, aug 2007.

\bibitem{pinkham}
R.~S. Pinkham.
\newblock On the distribution of first significant digits.
\newblock {\em The Annals of Mathematical Statistics}, 32(4):1223--1230, dec
  1961.

\bibitem{stanley2001}
P.~Bernaola-Galv{\'{a}}n, P.Ch. Ivanov, L.A.N. Amaral, and H.E. Stanley.
\newblock Scale invariance in the nonstationarity of human heart rate.
\newblock {\em Physical Review Letters}, 87(16), oct 2001.

\bibitem{fadel2004}
P.~J. Fadel, S.~M. Barman, S.~W. Phillips, and G.~L. Gebber.
\newblock Fractal fluctuations in human respiration.
\newblock {\em Journal of Applied Physiology}, 97(6):2056--2064, dec 2004.

\bibitem{cai2007}
S.~Cai, Z.~Jiang, T.~Zhou, P.~Zhou, H.~Yang, and Bing-Hong Wang.
\newblock Scale invariance of human electroencephalogram signals in sleep.
\newblock {\em Physical Review E}, 76(6), dec 2007.

\bibitem{gisiger2000}
T.~Gisiger.
\newblock Scale invariance in biology: coincidence or footprint of a universal
  mechanism?
\newblock {\em Biological Reviews of the Cambridge Philosophical Society},
  76(2):161--209, may 2001.

\bibitem{hill95}
T.~P. Hill.
\newblock A statistical derivation of the significant-digit law.
\newblock {\em Statistical Science}, 10(4):354--363, nov 1995.

\bibitem{li2015}
Q.~Liand~Z. Fu and N.~Yuan.
\newblock Beyond benford's law: Distinguishing noise from chaos.
\newblock {\em PloS one}, 10(6):e0129161, 2015.

\bibitem{Snyder2001}
M.A. Snyder, J.H. Curry, and A.M. Dougherty.
\newblock Stochastic aspects of one-dimensional discrete dynamical systems:
  Benford's law.
\newblock {\em Physical Review E}, 64(2), jul 2001.

\bibitem{Rodriguez2004}
R.J. Rodriguez.
\newblock Reducing false alarms in the detection of human influence on data.
\newblock {\em Journal of Accounting, Auditing \& Finance}, 19(2):141--158, apr
  2004.

\bibitem{millerweibullCuff2015}
V.Cuff, A.Lewis, and S.J. Miller.
\newblock The weibull distribution and benford's law.
\newblock {\em Involve, a Journal of Mathematics}, 8(5):859--874, sep 2015.

\bibitem{burke}
J.~Burke and E.~Kincanon.
\newblock Benford's law and physical constants: The distribution of initial
  digits.
\newblock {\em American Journal of Physics}, 59(10):952--952, oct 1991.

\bibitem{costas}
E.~Costas, V.~L{\'{o}}pez-Rodas, F.~Javier Toro, and A.~Flores-Moya.
\newblock The number of cells in colonies of the cyanobacterium microcystis
  aeruginosa satisfies benford{\textquotesingle}s law.
\newblock {\em Aquatic Botany}, 89(3):341--343, oct 2008.

\bibitem{buck}
B.~Buck, A.C. Merchant, and S.M. Perez.
\newblock An illustration of benford{\textquotesingle}s first digit law using
  alpha decay half lives.
\newblock {\em European Journal of Physics}, 14(2):59--63, mar 1993.

\bibitem{nigrini}
M.J. Nigrini.
\newblock {\em Benford{\textquotesingle}s Law}.
\newblock John Wiley {\&} Sons, Inc., jan 2012.

\bibitem{sinha2015}
P.~Seenivasan, S.~Easwaran, S.~Sridhar, and S.~Sinha.
\newblock Using skewness and the first-digit phenomenon to identify dynamical
  transitions in cardiac models.
\newblock {\em Frontiers in Physiology}, 6, jan 2016.

\bibitem{kreuzer2014}
M.~Kreuzer, D.~Jordan, B.~Antkowiak, B.~Drexler, E.~F. Kochs, and G.~Schneider.
\newblock Brain electrical activity obeys benford's law.
\newblock {\em Anesthesia {\&} Analgesia}, 118(1):183--191, jan 2014.

\bibitem{rsoftware}
{R Core Team}.
\newblock {\em R: A Language and Environment for Statistical Computing}.
\newblock R Foundation for Statistical Computing, Vienna, Austria, 2013.

\bibitem{hubbard1961}
J.I. Hubbard.
\newblock The effect of calcium and magnesium on the spontaneous release of
  transmitter from mammalian motor nerve endings.
\newblock {\em The Journal of Physiology}, 159(3):507--517, dec 1961.

\bibitem{pietronero}
L.~Pietronero, E.~Tosatti, V.~Tosatti, and A.~Vespignani.
\newblock Explaining the uneven distribution of numbers in nature: the laws of
  benford and zipf.
\newblock {\em Physica A: Statistical Mechanics and its Applications},
  293(1-2):297--304, apr 2001.

\bibitem{fu2007}
D.~Fu, Y.~Q. Shi, and W.~Su.
\newblock A generalized benford{\textquotesingle}s law for {JPEG} coefficients
  and its applications in image forensics.
\newblock In {\em Security, Steganography, and Watermarking of Multimedia
  Contents {IX}}, volume 6505. {SPIE}, feb 2007.

\bibitem{gauvrit2009}
N.~Gauvrit, J.C. Houillon, and J.P. Delahaye.
\newblock Generalized benford's law as a lie detector.
\newblock {\em Advances in Cognitive Psychology}, 13(2):121--127, jun 2017.

\bibitem{kossovsky}
A.E. Kossovsky.
\newblock {\em Benford's Law: Theory, the General Law of Relative Quantities,
  and Forensic Fraud Detection Applications}.
\newblock World Scientific Pub. Co. Inc., 2014.

\bibitem{slepkov2015}
A.D. Slepkov, K.B. Ironside, and D.~DiBattista.
\newblock Benford's law: Textbook exercises and multiple-choice testbanks.
\newblock {\em {PLOS} {ONE}}, 10(2):e0117972, feb 2015.

\bibitem{Greene2017}
W.H. Greene.
\newblock {\em Econometric Analysis}.
\newblock Pearson, 2017.

\bibitem{Myung2004}
J.I. Myung and M.A. Pitt.
\newblock Model comparison methods.
\newblock In {\em Methods in Enzymology}, pages 351--366. Elsevier, 2004.

\bibitem{Rahamimoff1978}
R.~Rahamimoff, S.D. Erulkar, A.~Lev-Tov, and H.~Meiri.
\newblock Intracellullar and extracellular calcium ions in transmitter release
  at the neuromuscular synapse.
\newblock {\em Annals of the New York Academy of Sciences}, 307(1):583--598,
  apr 1978.

\bibitem{muniak1982}
C.G. Carlson, M.E. Kriebel, and C.G. Muniak.
\newblock The effect of temperature on the amplitude distributions of miniature
  endplate potentials in the mouse diaphragm.
\newblock {\em Neuroscience}, 7(10):2537--2549, oct 1982.

\bibitem{shao}
L.~Shao and Bo-Qiang Ma.
\newblock First-digit law in nonextensive statistics.
\newblock {\em Physical Review E}, 82(4), oct 2010.

\bibitem{formam}
A.K. Formann.
\newblock The newcomb-benford law in its relation to some common distributions.
\newblock {\em {PLoS} {ONE}}, 5(5):e10541, may 2010.

\bibitem{Li2011}
C.~Li.
\newblock A model of neuronal intrinsic plasticity.
\newblock {\em {IEEE} Transactions on Autonomous Mental Development},
  3(4):277--284, dec 2011.

\bibitem{Park2014}
I.M. Park, Y.V. Bobkov, B.W. Ache, and J.C. Pr{\'{\i}}ncipe.
\newblock Intermittency coding in the primary olfactory system: A neural
  substrate for olfactory scene analysis.
\newblock {\em The Journal of Neuroscience}, 34(3):941--952, jan 2014.

\bibitem{Camproux1996}
A.C. Camproux, F.~Saunier, G.~Chouvet, J.C. Thalabard, and G.~Thomas.
\newblock A hidden markov model approach to neuron firing patterns.
\newblock {\em Biophysical Journal}, 71(5):2404--2412, nov 1996.

\bibitem{Bormashenko2016}
E.~Bormashenko, E.~Shulzinger, G.~Whyman, and Y.~Bormashenko.
\newblock Benford's law, its applicability and breakdown in the {IR} spectra of
  polymers.
\newblock {\em Physica A: Statistical Mechanics and its Applications},
  444:524--529, feb 2016.

\bibitem{lemons1985}
D.~S. Lemons.
\newblock On the numbers of things and the distribution of first digits.
\newblock {\em American Journal of Physics}, 54(9):816--817, sep 1986.

\bibitem{branco}
T.~Branco and K.~Staras.
\newblock The probability of neurotransmitter release: variability and feedback
  control at single synapses.
\newblock {\em Nature Reviews Neuroscience}, 10(5):373--383, may 2009.

\end{thebibliography}

\appendix
\section*{}



\section*{Supplementary material}

# Supplementary Information
## On the validation of Newcomb-Benford law and Weibull distribution in neuromuscular transmission

A. J. da Silva

Instituto de Humanidades, Artes e Ciências
Universidade Federal do Sul da Bahia
Itabuna, Bahia.Brazil
adjesbr@ufsb.edu.br,adjesbr@gmail.com

S. Floquet

Colegiado de Engenharia Civil
Universidade Federal do Vale do São Francisco
Juazeiro, Bahia, Brazil

D. O. C. Santos

Instituto de Humanidades, Artes e Ciências
Universidade Federal do Sul da Bahia
Itabuna, Bahia. Brazil

R. F. Lima

Departamento de Fisiologia e Farmacologia
Faculdade de Medicina, Universidade Federal do Ceará
Fortaleza, Ceará, Brazil

October 22, 2019

The following table gives results regarding ample sizes, model selection criteria values (AIC - Akaike Information Criteria and BIC - Bayesian Information Criteria). It gives also adjusted model parameter values for the best fitted distribution.

Table 1: Statistical summary of data used in the paper.

| Concentration (mM) | Sample Size | AIC Weibull | AIC lognormal | BIC Weibull | BIC lognormal | Best Distribution | Best Distribution Parameters |
|---|---|---|---|---|---|---|---|
| 0.6 | 463 | 8354.20 | 8428.67 | 8362.47 | 8436.95 | Weibull | r=3125.813 b=1.070 |
| 0.6 | 213 | 3879.31 | 3886.06 | 3886.03 | 3892.79 | Weibull | r=3330.704 b=1.031 |
| 0.6 | 252 | 4501.81 | 4551.81 | 4508.87 | 4558.87 | Weibull | r=2772.104 b=1.008 |

1

Table 1: Statistical summary of data used in the paper.

| Concentration (mM) | Sample Size | AIC Weibull | AIC lognormal | BIC Weibull | BIC lognormal | Best Distribution | Best Distribution Parameters |
|---|---|---|---|---|---|---|---|
| 0.6 | 289 | 5092.28 | 5145.52 | 5099.61 | 5152.85 | Weibull | r=2512.852 b=1.060 |
| 0.6 | 571 | 10471.90 | 10537.06 | 10480.59 | 10545.75 | Weibull | r=3357.899 b=0.889 |
| 0.6 | 1620 | 29526.83 | 29747.62 | 29537.61 | 29758.40 | Weibull | r=3339.153 b=1.004 |
| 0.6 | 1494 | 25364.14 | 25534.81 | 25374.76 | 25545.42 | Weibull | r=1788.477 b=1.004 |
| 0.6 | 681 | 12533.68 | 12578.14 | 12542.73 | 12587.19 | Weibull | r=3469.084 b=0.892 |
| 0.6 | 1019 | 17824.60 | 17940.18 | 17834.45 | 17950.04 | Weibull | r=2328.031 b=1.020 |
| 0.6 | 624 | 11605.64 | 11704.70 | 11614.51 | 11713.57 | Weibull | r=4060.891 b=1.030 |
| 0.6 | 1621 | 28769.18 | 29018.18 | 28779.96 | 29028.96 | Weibull | r=2637.961 b=1.012 |
| 0.6 | 2432 | 42351.50 | 42716.07 | 42363.09 | 42727.67 | Weibull | r=2237.991 b=1.017 |
| 0.6 | 2471 | 42945.34 | 43223.34 | 42956.97 | 43234.96 | Weibull | r=2170.007 b=0.985 |
| 0.6 | 1050 | 20045.18 | 20223.53 | 20055.09 | 20233.45 | Weibull | r=5129.077 b=0.998 |
| 1.2 | 775 | 12933.46 | 13006.33 | 12942.76 | 13015.63 | Weibull | r=1561.943 b=1.028 |
| 1.2 | 585 | 10064.30 | 10104.34 | 10073.04 | 10113.09 | Weibull | r=1956.114 b=0.955 |
| 1.2 | 244 | 4553.50 | 4595.41 | 4560.50 | 4602.41 | Weibull | r=4251.110 b=1.076 |
| 1.2 | 473 | 7974.68 | 7937.22 | 7983.00 | 7945.53 | lognormal | $\mu$=6.648 $\sigma$=1.378 |
| 1.2 | 1835 | 32990.05 | 33236.22 | 33001.08 | 33247.25 | Weibull | r=2959.899 b=1.011 |
| 1.2 | 2472 | 42957.54 | 43301.31 | 42969.17 | 43312.94 | Weibull | r=2198.702 b=1.018 |
| 1.2 | 3468 | 57915.47 | 58369.68 | 57927.77 | 58381.98 | Weibull | r=1572.010 b=1.025 |
| 1.2 | 2632 | 45410.54 | 45778.68 | 45422.29 | 45790.43 | Weibull | r=2044.036 b=0.992 |
| 1.2 | 978 | 18056.48 | 18194.19 | 18066.25 | 18203.96 | Weibull | r=3741.418 b=0.995 |
| 1.2 | 2369 | 40983.87 | 41328.65 | 40995.41 | 41340.19 | Weibull | r=2115.485 b=1.018 |
| 1.2 | 1195 | 21235.47 | 21437.29 | 21245.64 | 21447.47 | Weibull | r=2602.688 b=0.954 |
| 1.8 | 730 | 11882.86 | 11948.62 | 11892.04 | 11957.81 | Weibull | r=1245.394 b=0.979 |
| 1.8 | 869 | 13905.27 | 14020.06 | 13914.80 | 14029.60 | Weibull | r=1123.562 b=1.060 |
| 1.8 | 2181 | 30673.86 | 30776.52 | 30685.24 | 30787.89 | Weibull | r=869.668 b=0.978 |
| 1.8 | 928 | 14439.41 | 14501.21 | 14449.08 | 14510.88 | Weibull | r=869.668 b=0.978 |
| 1.8 | 2973 | 41789.38 | 41929.36 | 41801.37 | 41941.35 | Weibull | r=432.438 b=1.094 |
| 1.8 | 642 | 10976.21 | 11056.26 | 10985.14 | 11065.19 | Weibull | r=1910.256 b=1.022 |
| 1.8 | 1349 | 23613.66 | 23719.39 | 23624.07 | 23729.80 | Weibull | r=2308.445 b=0.985 |
| 1.8 | 1162 | 19750.02 | 19865.06 | 19760.14 | 19875.18 | Weibull | r=1797.833 b=0.995 |
| 1.8 | 1685 | 29734.03 | 29953.06 | 29744.89 | 29963.92 | Weibull | r=2461.072 b=0.968 |
| 1.8 | 1009 | 18263.11 | 18362.41 | 18272.95 | 18372.24 | Weibull | r=3160.157 b=1.024 |
| 1.8 | 2048 | 35343.02 | 35565.73 | 35354.27 | 35576.98 | Weibull | r=2074.024 b=1.022 |
| 1.8 | 3060 | 50346.97 | 50673.32 | 50359.02 | 50685.38 | Weibull | r=1385.304 b=1.018 |
| 1.8 | 1510 | 26980.77 | 27147.15 | 26991.41 | 27157.79 | Weibull | r=2798.198 b=1.009 |
| 1.8 | 2436 | 41190.27 | 41449.69 | 41201.87 | 41461.28 | Weibull | r=1732.496 b=1.008 |
| 2.4 | 602 | 10709.15 | 10798.61 | 10717.95 | 10807.41 | Weibull | r=2714.169 b=1.035 |
| 2.4 | 498 | 8357.89 | 8402.87 | 8366.31 | 8411.29 | Weibull | r=1624.869 b=1.013 |
| 2.4 | 624 | 10019.69 | 10082.47 | 10028.57 | 10091.35 | Weibull | r=1134.855 b=1.020 |
| 2.4 | 671 | 11654.46 | 11733.78 | 11663.47 | 11742.80 | Weibull | r=2209.671 b=1.044 |
| 2.4 | 1037 | 15891.29 | 15995.06 | 15901.18 | 16004.95 | Weibull | r=799.759 b=1.055 |
| 2.4 | 883 | 13977.95 | 14036.58 | 13987.51 | 14046.14 | Weibull | r=1020.370 b=1.034 |

2

Table 1: Statistical summary of data used in the paper.

| Concentration (mM) | Sample Size | AIC Weibull | AIC lognormal | BIC Weibull | BIC lognormal | Best Distribution | Best Distribution Parameters |
|---|---|---|---|---|---|---|---|
| 2.4 | 1138 | 16833.20 | 16903.86 | 16843.27 | 16913.93 | Weibull | r=615.306 b=1.063 |
| 4.8 | 1008 | 16735.31 | 16853.06 | 16745.14 | 16862.90 | Weibull | r=1525.433 b=1.071 |
| 4.8 | 910 | 14376.92 | 14471.08 | 14386.55 | 14480.70 | Weibull | r=1026.688 b=1.085 |
| 4.8 | 741 | 11871.27 | 11931.32 | 11880.49 | 11940.54 | Weibull | r=1107.344 b=1.005 |
| 4.8 | 1864 | 27586.85 | 27690.66 | 27597.91 | 27701.72 | Weibull | r=609.204 b=1.030 |
| 4.8 | 1021 | 16084.53 | 16170.36 | 16094.39 | 16180.21 | Weibull | r=964.116 b=0.991 |
| 4.8 | 1116 | 17199.21 | 17293.53 | 17209.25 | 17303.56 | Weibull | r=825.880 b=1.029 |
| 4.8 | 1122 | 17480.60 | 17598.72 | 17490.65 | 17608.76 | Weibull | r=913.554 b=1.067 |
| 10.0 | 2757 | 41243.63 | 41506.01 | 41255.47 | 41517.86 | Weibull | r=668.122 b=1.058 |
| 10.0 | 1824 | 27574.87 | 27751.12 | 27585.89 | 27762.14 | Weibull | r=705.565 b=1.003 |
| 10.0 | 2048 | 32842.47 | 33024.05 | 32853.72 | 33035.30 | Weibull | r=1122.269 b=1.014 |
| 10.0 | 3724 | 55012.21 | 55228.03 | 55024.66 | 55240.47 | Weibull | r=603.834 b=1.040 |
| 10.0 | 2991 | 44266.03 | 44481.18 | 44278.04 | 44493.19 | Weibull | r=605.018 b=1.014 |
| 10.0 | 2809 | 41915.24 | 42163.77 | 41927.12 | 42175.65 | Weibull | r=653.808 b=1.052 |
| 10.0 | 2819 | 42048.05 | 42204.54 | 42059.94 | 42216.43 | Weibull | r=650.601 b=1.047 |
| 10.0 | 2178 | 33611.51 | 33759.74 | 33622.88 | 33771.11 | Weibull | r=839.966 b=1.041 |
| 10.0 | 1228 | 20352.49 | 20521.83 | 20362.71 | 20532.06 | Weibull | r=1480.594 b=1.035 |
| 10.0 | 1511 | 24425.58 | 24576.64 | 24436.22 | 24587.28 | Weibull | r=1201.893 b=1.024 |
| 10.0 | 2342 | 35473.26 | 35676.76 | 35484.78 | 35688.28 | Weibull | r=733.248 b=1.057 |
| 10.0 | 2216 | 34118.73 | 34306.74 | 34130.13 | 34318.14 | Weibull | r=825.766 b=1.043 |
| 10.0 | 2703 | 40540.11 | 40687.59 | 40551.91 | 40699.40 | Weibull | r=673.718 b=1.033 |
| 15.0 | 3213 | 42608.54 | 42701.69 | 42620.69 | 42713.84 | Weibull | r=289.022 b=1.082 |
| 15.0 | 2242 | 31343.30 | 31463.93 | 31354.73 | 31475.36 | Weibull | r=414.211 b=1.084 |
| 15.0 | 2294 | 33283.42 | 33296.53 | 33294.89 | 33308.01 | Weibull | r=539.385 b=1.081 |
| 15.0 | 2328 | 35617.21 | 35820.59 | 35628.71 | 35832.09 | Weibull | r=780.920 b=1.027 |
| 15.0 | 445 | 7299.81 | 7358.86 | 7308.01 | 7367.06 | Weibull | r=1356.476 b=1.035 |
| 15.0 | 1925 | 30183.29 | 30332.90 | 30194.42 | 30344.02 | Weibull | r=942.651 b=1.023 |
| 15.0 | 3464 | 41886.88 | 41892.97 | 41899.19 | 41905.27 | Weibull | r=159.048 b=1.052 |
| 15.0 | 2277 | 33071.01 | 33135.13 | 33082.47 | 33146.59 | Weibull | r=506.781 b=0.925 |
| 15.0 | 4123 | 52549.98 | 52154.28 | 52562.63 | 52166.92 | lognormal | $\mu$=4.723 $\sigma$=1.201 |
| 15.0 | 1064 | 17080.36 | 17137.50 | 17090.30 | 17147.44 | Weibull | r=1101.935 b=0.956 |
| 15.0 | 1517 | 24585.73 | 24659.27 | 24596.38 | 24669.92 | Weibull | r=1179.773 b=0.935 |
| 15.0 | 1880 | 29561.57 | 29784.76 | 29572.65 | 29795.84 | Weibull | r=970.410 b=1.039 |

3



\end{document}